\documentclass[times,twocolumn,final,longtitle,nopreprintline]{elsarticle}

\usepackage{medima}
\PassOptionsToPackage{bookmarks=false}{hyperref}
\usepackage[utf8]{inputenc}
\usepackage{xcolor}
\usepackage{colortbl}
\usepackage{float}
\usepackage{mathrsfs}
\usepackage{tikz}
\usepackage{comment}
\usepackage{dblfloatfix}
\usepackage{enumitem}
\usepackage{multirow}
\usepackage{makecell}
\usepackage{tablefootnote}
\usepackage{booktabs}
\usepackage[bookmarks=false]{hyperref}
\hypersetup{
    colorlinks=false,
     pdfborder={0 0 0},
}

\definecolor{OvGUMEMoRIAL}{HTML}{F7DB2B}
\definecolor{ISDUE}{HTML}{475BF9}
\definecolor{lachinov}{HTML}{347AFD}
\definecolor{IITKGP-KLIV}{HTML}{463DE0}
\definecolor{MedianCHAOS1}{HTML}{2797EB}
\definecolor{MedianCHAOS2}{HTML}{17AEDA}
\definecolor{MedianCHAOS3}{HTML}{12BEB9}
\definecolor{MedianCHAOS4}{HTML}{3DC990}
\definecolor{MedianCHAOS5}{HTML}{81CC59}
\definecolor{MedianCHAOS6}{HTML}{C8C129}
\definecolor{PKDIA}{HTML}{F9FB15}
\definecolor{mountain}{HTML}{FCBB3E}
\definecolor{CIR-MPerkonigg}{HTML}{3E26A8}
\definecolor{PKDIAv2}{HTML}{F9FB15}
\definecolor{METU_MMLAB}{HTML}{FF0000}
\definecolor{nnU-Net}{HTML}{F72A9B}
\newcommand{\teamcolor}[1]{
\begin{tikzpicture}
\draw [fill=#1, draw=#1]circle (1.0mm);
\end{tikzpicture}}
\usepackage{soul}
\newcommand{\lk}[1]{\textcolor{black}{#1}}
\newcommand{\emre}[1]{\textcolor{black}{#1}}

\newcommand{\quot}[1]{``#1''}

\newcolumntype{L}[1]{>{\raggedright\let\newline\\\arraybackslash\hspace{0pt}}m{#1}}
\newcolumntype{C}[1]{>{\centering\let\newline\\\arraybackslash\hspace{0pt}}m{#1}}
\newcolumntype{R}[1]{>{\raggedleft\let\newline\\\arraybackslash\hspace{0pt}}m{#1}}

\journal{Medical Image Analysis}
\begin{document}
\verso{A.E. Kavur \textit{et~al.}}
\begin{frontmatter}
\title{CHAOS challenge - combined (CT-MR) Healthy Abdominal Organ Segmentation}%

\author[1]{A. Emre \snm{Kavur}\corref{cor1}}
\cortext[cor1]{Corresponding author: 
  e-mail:~emrekavur@gmail.com}  
\author[2]{N.~Sinem \snm{Gezer}}
\author[2]{Mustafa \snm{Barış}}
\author[15,22]{Sinem \snm{Aslan}}
\author[4]{Pierre-Henri \snm{Conze}} 
\author[5]{Vladimir \snm{Groza}} 
\author[6]{Duc~Duy \snm{Pham}} 
\author[7,21]{Soumick \snm{Chatterjee}} 
\author[7]{Philipp \snm{Ernst}} 
\author[8]{Savaş \snm{Özkan}} 
\author[8]{Bora \snm{Baydar}} 
\author[9]{Dmitry \snm{Lachinov}} 
\author[10]{Shuo \snm{Han}} 
\author[6]{Josef \snm{Pauli}}  
\author[11]{Fabian \snm{Isensee}} 
\author[12]{Matthias \snm{Perkonigg}} 
\author[13]{Rachana \snm{Sathish}} 
\author[14]{Ronnie \snm{Rajan}} 
\author[13]{Debdoot \snm{Sheet}} 
\author[6]{Gurbandurdy \snm{Dovletov}}  
\author[21]{Oliver \snm{Speck}} 
\author[7]{Andreas \snm{Nürnberger}} 
\author[11]{Klaus~H. \snm{Maier-Hein}}  
\author[8]{Gözde \snm{Bozdağı~Akar}} 
\author[16]{Gözde \snm{Ünal}}
\author[2]{Oğuz \snm{Dicle}}
\author[17]{M.~Alper \snm{Selver}\corref{cor2}}
\cortext[cor2]{Corresponding author: 
	e-mail:~alper.selver@deu.edu.tr} 

\address[1]{Graduate School of Natural and Applied Sciences, Dokuz Eylul University, Izmir, Turkey}
\address[2]{Department of Radiology, Faculty Of Medicine, Dokuz Eylul University, Izmir, Turkey}
\address[15]{Ca' Foscari University of Venice, ECLT and DAIS, Venice, Italy}%
\address[22]{Ege University, International Computer Institute, Izmir, Turkey}%
\address[4]{IMT Atlantique, LaTIM UMR 1101, Brest, France }%
\address[5]{Median Technologies, Valbonne, France}%
\address[6]{Intelligent Systems, Faculty of Engineering, University of Duisburg-Essen, Germany}%
\address[7]{Data and Knowledge Engineering Group, Otto von Guericke University, Magdeburg, Germany}%
\address[21]{Biomedical Magnetic Resonance, Otto von Guericke University Magdeburg, Germany.}%
\address[8]{Department of Electrical and Electronics Engineering, Middle East Technical University, Ankara, Turkey}%
\address[9]{Department of Ophthalmology and Optometry, Medical Uni. of Vienna, Austria}%
\address[10]{Johns Hopkins University, Baltimore, USA}%
\address[11]{Division of Medical Image Computing, German Cancer Research Center, Heidelberg,~Germany}%
\address[12]{CIR Lab Dept of Biomedical Imaging and Image-guided Therapy Medical Uni. of Vienna, Austria}%
\address[13]{Department of Electrical Engineering, Indian Institute of Technology, Kharagpur, India}%
\address[14]{School of Medical Science and Technology, Indian Institute of Technology, Kharagpur, India}%
\address[16]{Department of Computer Engineering, İstanbul Technical University, İstanbul, Turkey }%
\address[17]{Department of Electrical and Electronics Engineering, Dokuz Eylul University, Izmir, Turkey}

\received{-}
\finalform{-}
\accepted{-}
\availableonline{-}
\communicated{-}

\begin{abstract} 
\small
Segmentation of abdominal organs has been a comprehensive, yet unresolved, research field for many years. In the last decade, intensive developments in deep learning (DL) introduced new state-of-the-art segmentation systems. Despite outperforming the overall accuracy of existing systems, the effects of DL model properties and parameters on the performance are hard to interpret. This makes comparative analysis a necessary tool \lk{towards interpretable} 
studies and systems. Moreover, the performance of DL for emerging learning approaches such as cross-modality and multi-modal semantic segmentation tasks has been rarely discussed. In order to expand the knowledge on these topics, the CHAOS~--~Combined (CT-MR) Healthy Abdominal Organ Segmentation challenge \emre{was} organized in conjunction with the IEEE International Symposium on Biomedical Imaging (ISBI), 2019, in Venice, Italy. Abdominal organ segmentation from routine acquisitions plays an \emre{important} role in several clinical applications, such as pre-surgical planning or morphological and volumetric follow-ups for various diseases. These applications require a certain level of performance on a diverse set of metrics such as maximum symmetric surface distance (MSSD) to determine surgical error-margin or overlap errors for tracking size and shape differences. Previous abdomen related challenges are mainly focused on tumor/lesion detection and/or classification with a single modality. Conversely, CHAOS provides both abdominal CT and MR data from healthy subjects for single and multiple abdominal organ segmentation. Five different but complementary tasks \emre{were} designed to analyze the capabilities of participating approaches from multiple perspectives. The results \emre{were} investigated thoroughly, compared with manual annotations and interactive methods. The analysis shows that the performance of DL models for single modality (CT / MR) can show reliable volumetric analysis performance (DICE: 0.98 $\pm$ 0.00 / 0.95 $\pm$ 0.01), but the best MSSD performance remains limited (21.89 $\pm$ 13.94 / 20.85 $\pm$ 10.63 mm). The performances of participating models decrease \emre{dramatically} for cross-modality tasks both for the liver (DICE: 0.88 $\pm$ 0.15 MSSD: 36.33 $\pm$ 21.97 mm). Despite contrary examples on different applications, multi-tasking DL models designed to segment all organs are observed to perform worse compared to organ-specific ones (performance drop around 5\%). Nevertheless, some of the successful models show better performance with their multi-organ versions. We conclude that the exploration of those pros and cons in both single vs multi-organ and cross-modality segmentations is poised to have an impact on further research for developing effective algorithms that would support real-world clinical applications. Finally, having more than 1500 participants and receiving more than 550 submissions, another important contribution of this study is the analysis on shortcomings of challenge organizations such as the effects of multiple submissions and peeking \lk{phenomenon}.
\end{abstract}

\begin{keyword}
\KWD Segmentation\sep Challenge\sep Abdomen\sep Cross-modality
\end{keyword}
\end{frontmatter}

\section{Introduction}
In the last decade, medical imaging and image processing benchmarks have become effective venues to compare performance of different approaches in clinically important tasks \citep{ayache201620th}. These benchmarks have gained a particularly important role in the analysis of learning-based systems by enabling the use of common datasets for training and testing \citep{simpson2019large}. Challenges \emre{that} use these benchmarks, bear a prominent role in reporting outcomes of the state-of-the-art results in a structured way \citep{kozubek2016challenges}. In this respect, the benchmarks establish standard datasets, evaluation strategies, fusion possibilities (e.g. ensembles), and \emre{unresolved} difficulties related to the specific biomedical image processing task(s) being tested \citep{menze2014multimodal}. An extensive website, grand-challenge.org \citep{vanGinneken2015}, has been designed for hosting the challenges related to medical image segmentation and currently includes around 200 challenges. 

A comprehensive exploration of biomedical image analysis challenges reveals that the construction of datasets, inter- and intra-observer variations for ground truth generation as well as evaluation criteria might prevent establishing the true potential of such events \citep{reinke2018winner}. Suggestions, caveats, and roadmaps are being provided by reviews~\citep{Maier-Hein2018,reinke2018exploit} to improve the challenges. 

Considering the dominance of machine learning (ML) approaches, two main points are continuously being emphasized: 1) recognition of current roadblocks in applying ML to medical imaging, 2) increasing the dialogue between radiologists and data scientists \citep{prevedello2019challenges}. Accordingly, challenges are either continuously updated \citep{menze2014multimodal}, repeated after some time \citep{staal2004ridge}, or new ones having similar focuses are being organized to overcome the pitfalls and shortcomings of existing ones. 

Abdominal imaging is one of the important sub-fields of diagnostic radiology. It focuses on imaging the organs/structures in the abdomen such as the liver, kidneys, spleen, bladder, prostate, pancreas by CT, MRI, ultrasonography, or any other dedicated imaging modality. Emergencies that require treatment or intervention such as acute liver failure, impaired kidney function, and abdominal aortic aneurysm must be immediately detected by abdominal imaging. It plays important role in identifying various diseases during routine controls and follow-ups. Therefore, studies and challenges in the segmentation of abdominal organs/structures have always constituted an important research field.

A detailed literature review about the challenges related to abdominal organs (see Section II) revealed that the existing challenges in the field are dominated by CT scans and tumor/lesion classification tasks. Up to now, there have only been a few benchmarks containing abdominal MRI series (Table I). Although this situation was typical for the last decades, the emerging technology of MRI makes it the preferred modality for further and detailed analysis of the abdomen. The remarkable developments in MRI technology in terms of resolution, dynamic range, and speed enable joint analyses of these modalities \citep{hirokawa2008mri}.


To gauge the current state-of-the-art in automated abdominal segmentation and observe the performance of various approaches on different tasks such as cross-modality learning and multi-modal segmentation, we organized the Combined (CT-MR) Healthy Abdominal Organ Segmentation (CHAOS) challenge in conjunction with the IEEE International Symposium on Biomedical Imaging (ISBI) in 2019. For this purpose, we prepared and made available a unique dataset of CT and MR scans from unpaired abdominal image series. A consensus-based multiple expert annotation strategy was used to generate ground truth. A subset of this dataset was provided to the participants for training, and the remaining images were used to test performance against the (hidden) manual delineations using various metrics. In this paper, we report both setup and the results of this CHAOS benchmark as well as its outcomes.

The rest of the paper is organized as follows. A review of the current challenges in abdominal organ segmentation is given in Section II together with surveys on benchmark methods. Next, CHAOS datasets, setup, ground truth generation, and \lk{the }
tasks are presented in Section III. Section IV describes the evaluation strategy. Then, participating methods are comparatively summarized in Section V. Section VI presents the results, and Section VII provides \lk{a} discussion and concludes the paper.

\section{Related Work}

\begin{table*}[!t]
\def\arraystretch{1.1}
\caption{Overview of challenges that have upper abdomen data and task. (Other structures are not shown in the table.)}
\label{tab:challenges}
\begin{tabular}{L{3.5cm}L{3.0cm}L{6.0cm}L{4.0cm}}
\toprule 
Challenge & Task(s) & Structure~(Modality) & Organization and year\\ \midrule 
SLIVER07 \citep{van20073d} & Single model segmentation & Liver~(CT) & MICCAI 2007, Australia \\ \midrule
LTSC08 \newline \citep{deng20083d} & Single model segmentation & Liver tumor~(CT) & MICCAI 2008, USA \\ \midrule
Shape 2014 \newline \citep{kistler2013virtual}& Building organ model & Liver~(CT) & Delémont,  Switzerland \\ \midrule
Shape 2015 \newline \citep{kistler2013virtual}& Completing partial segmentation & Liver (CT) & Delémont, Switzerland \\ \midrule
VISCERAL Anatomy 3 \newline \citep{jimenez2016cloud}& Multi-model segmentation & Kidney, urinary bladder, gallbladder, spleen, liver, and pancreas~(CT and MRI for all organs) & VISCERAL Consortium, 2014 \\ \midrule
Multi-Atlas Labeling Beyond the Cranial Vault \newline \citep{multiatlas2015}& Multi-atlas segmentation & Adrenal glands, aorta, esophagus, gall bladder, kidneys, liver, pancreas, splenic/portal veins, spleen, stomach, and vena cava (CT) & MICCAI 2015 \\ \midrule
LiTS \newline \citep{bilic2019liver}& Single model segmentation & Liver and liver tumor~(CT) & ISBI~2017,~Australia; \newline MICCAI~2017,~Canada \\ \midrule
Pancreatic Cancer Survival Prediction \newline \citep{guinney2017prediction}& Quantitative assessment of cancer & Pancreas (CT) & MICCAI 2018, Spain \\ \midrule
MSD \newline \citep{simpson2019large}& Multi-model segmentation & Liver~(CT), liver tumor~(CT), spleen~(CT), hepatic vessels in the liver~(CT), pancreas and pancreas tumor~(CT)  & MICCAI 2018, Spain \\ \midrule
KiTS19 \newline \citep{KiTS19}& Single model segmentation & Kidney and kidney tumor~(CT) & MICCAI 2019, China \\ \midrule
CHAOS & Multi-model segmentation & Liver, kidney(s), spleen~(CT, MRI for all organs) & ISBI 2019, Italy \\
\bottomrule
\end{tabular}
\end{table*}

According to our literature analysis, currently, there exist 12 challenges focusing on abdominal organs \citep{vanGinneken2015} (see Tab. \ref{tab:challenges}). Being one of the pioneering challenges, SLIVER07 initialized the liver benchmarking \citep{ heimann2009comparison, van20073d}. It provided a comparative study of a range of algorithms for liver segmentation under several intentionally included difficulties such as patient orientation variations or tumors and lesions. Its outcomes reported a snapshot of the methods that were popular for medical image analysis at \lk{that} time. However, since then, abdomen-related challenges mostly targeted disease and tumor detection rather than organ segmentation. In 2008, “3D Liver Tumor Segmentation Challenge (LTSC08)” \citep{deng20083d} was organized as the continuation of the SLIVER07 challenge to segment liver tumors from abdomen CT scans. Similarly, Shape 2014 and 2015 \citep{kistler2013virtual} challenges focused on liver segmentation from CT data. VISCERAL Anatomy 3 \citep{jimenez2016cloud} provided a unique challenge, which was a very comprehensive platform for segmenting not only upper-abdominal organs, but also various \lk{other organs} such as left/right lung, urinary bladder, and pancreas. ``Multi-Atlas Labeling Beyond the Cranial Vault - Workshop and Challenge'' focused on multi-atlas segmentation with abdominal and cervix clinically acquired CT scans~\citep{multiatlas2015}. LiTS - Liver Tumor Segmentation Challenge \citep{bilic2019liver} is another example that covers liver and liver tumor segmentation tasks in CT. Other similar challenges can be listed as Pancreatic Cancer Survival Prediction \citep{guinney2017prediction}, which targets pancreas cancer tissues in CT scans; KiTS19 \citep{KiTS19} challenge, which provides CT data for kidney tumor segmentation.

In 2018, Medical Segmentation Decathlon (MSD) \citep{simpson2019large} was organized by a joint team and provided \lk{a substantial} 
challenge that contained many structures such as liver parenchyma, hepatic vessels and tumors, spleen, brain tumors, hippocampus, and lung tumors. The focus of the challenge was not only to evaluate the performance for each structure, but to observe the generalizability, translatability, and transferability of a system. Thus, the main idea behind MSD was to understand the key elements of DL systems that can work on many tasks. To provide such a source, MSD included a wide range of challenges including small and unbalanced sample sizes, varying object scales, and multi-class labels. The approach of MSD underlines the ultimate goal of the challenges that is to provide extensive datasets on several different tasks, and evaluation through a standardized analysis and validation process. 

In this respect, a recent survey showed that another trend in medical image segmentation is the development of more comprehensive computational anatomical models leading to multi-organ related tasks rather than traditional organ and/or disease-specific tasks \citep{CERROLAZA201944}. By incorporating inter-organ relations into the process, multi-organ related tasks require a complete representation of the complex and flexible abdominal anatomy. Thus, this emerging field requires new efficient computational and machine learning models.

\lk{Inspired by} the above-mentioned visionary studies, CHAOS \emre{was} organized to strengthen the field by aiming at objectives that involve emerging ML concepts, including cross-modality learning, and multi-modal segmentation. In this respect, \lk{CHAOS} focuses on segmenting multiple organs from unpaired patient datasets acquired by two modalities: CT and MR (including two different pulse sequences). 

\section{CHAOS Challenge}

\subsection{Data Information and Details}
The CHAOS challenge data contains 80 patients. 40 of them \emre{(22 male, 18 female, ages between 18 and 63 with average 44.85$\pm$11.29)} went through a single CT scan and 40 of them \emre{(23 male, 17 female, ages between 18 and 76 with average 54.60$\pm$14.25)} went through MR scans including 2 pulse sequences of the upper abdomen area. We present example images for CT and MR modalities in Fig.\ref{fig:samples}. Both CT and MR datasets include healthy abdomen organs without any pathological abnormalities (tumors, metastasis, and so on). 

There are various clinical \lk{reasons for measurement of} volume, size, and shape through the precise segmentation of healthy abdominal organs. For instance, the liver volume is affected by several diseases including congestive heart failure, cancer, cirrhosis, infections, metabolic disorders, and congenital diseases. The dimensions of the liver may give clues about the severity of the disease. The growth pattern of the liver and its change in the treatment process also provide valuable information about the importance and prognosis of the disease. For  this  reason,  determining  whether  the  liver  is  enlarged  or not,  calculating  its  volume,  and  specifying related effects are very important. \lk{Precise segmentation is also required to plan liver transplant surgeries. For example, determining whether a portion of the liver to be resected is sufficient for the recipient patient and whether the remaining liver will be sufficient for the donor is an important part of treatment decisions} \citep{Low2008}. Furthermore, the determination of the most suitable donors for living donated transplantation and pre-operative planning needs accurate segmentation of the liver. For these reasons, the objective was to evaluate the liver volume perfectly in the challenge, and accordingly, healthy patient livers were studied. Besides, there are several medical reasons that require the segmentation of not only the liver but also other solid organs in the abdominal region. For example, the spleen enlarges in cases of portal hypertension, infections and especially in lymphoproliferative diseases \citep{Robertson2001}. Because of its amorphous structure, a 3-dimensional visualization, which requires segmentation, can provide much better information about the organ compared to a 2-dimentional image analysis \citep{Joiner2015, Lamb2002, Linguraru2013}. Furthermore, segmentation can be used for monitoring the cortex thickness of kidneys, calculating cyst-parenchyma ratios as in polycystic kidney disease and particularly valuable for volumetric monitoring in renal tumors \citep{KING1505}.

The datasets \lk{for the CHAOS challenge} were collected from the Department of Radiology, Dokuz Eylul University Hospital, Izmir, Turkey. The scan protocols are briefly explained in the following subsections. Further details and explanations are available on the CHAOS website\footnote{CHAOS data information: https://chaos.grand-challenge.org/Data/}. This  study  was  approved  by  the  Institutional  Review  Board of Dokuz Eylul University.

\subsubsection{CT Data Specifications}
\begin{figure*}[!t]
    \centering
        \includegraphics[width=1.0\textwidth]{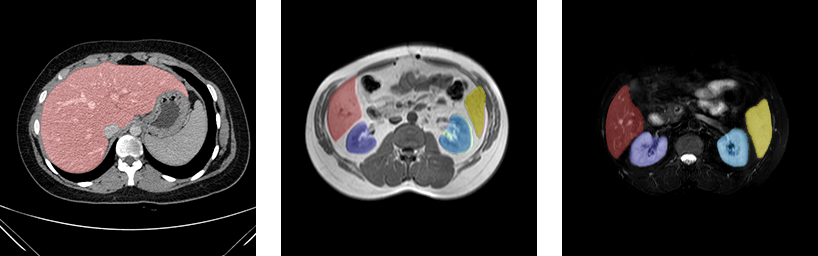}
        \caption{Example slices from CHAOS CT, MR (T1-DUAL in-phase) and MR (T2-SPIR) datasets (liver:red, right kidney:dark blue, left kidney:light blue and spleen:yellow).}
\label{fig:samples} 
\end{figure*}

\begin{figure}[!]
    \centering
        \includegraphics[width=0.48\textwidth]{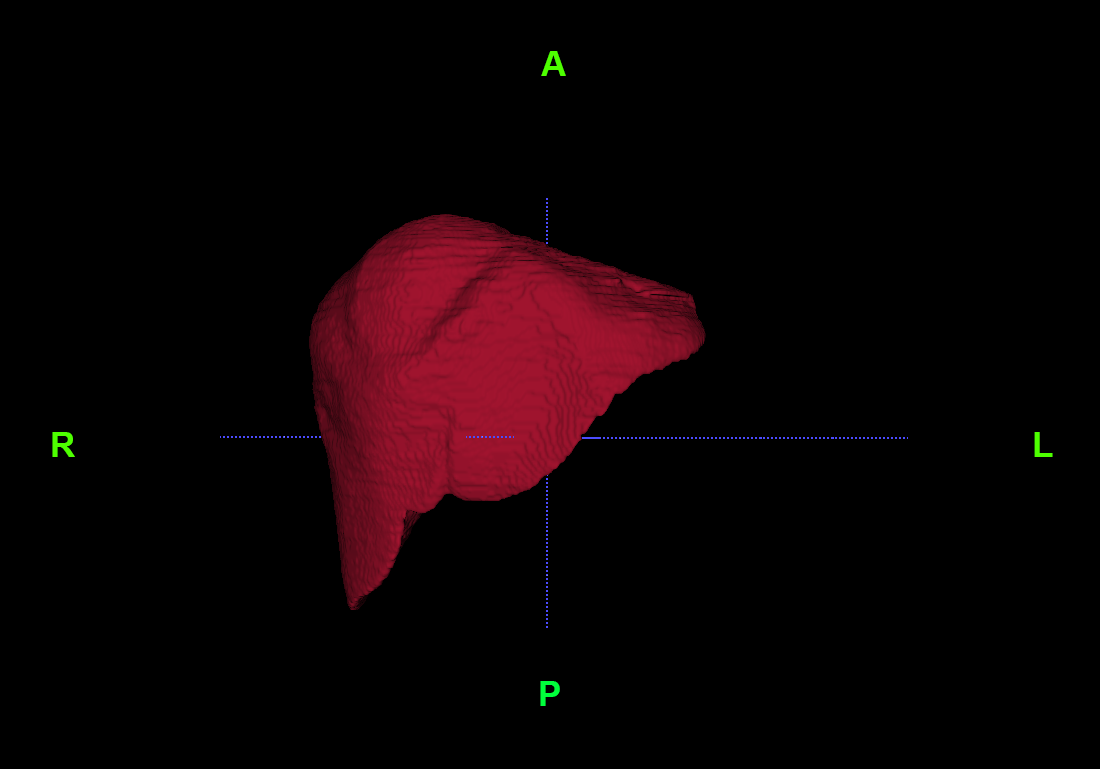}
        \caption{3D visualization of the liver from the CHAOS CT dataset (case 35).}
\label{fig:ct_sample} 
\vspace{3mm}
\end{figure}

The CT volumes were acquired at the portal venous phase after contrast agent injection. In this phase, the liver parenchyma is enhanced maximally through blood supply by the portal vein. Portal veins are enhanced \lk{well} but some enhancements also exist for hepatic veins. This phase is widely used for liver and vessel segmentation, prior to surgery. Since the tasks related to CT data only include liver segmentation, this set has only annotations for the liver. The details of the data are presented in Tab.\ref{tab:data_stats} and a sample case is illustrated in Fig.\ref{fig:samples}, left and Fig.\ref{fig:ct_sample}.


\begin{figure}[!]
    \centering
        \includegraphics[width=0.48\textwidth]{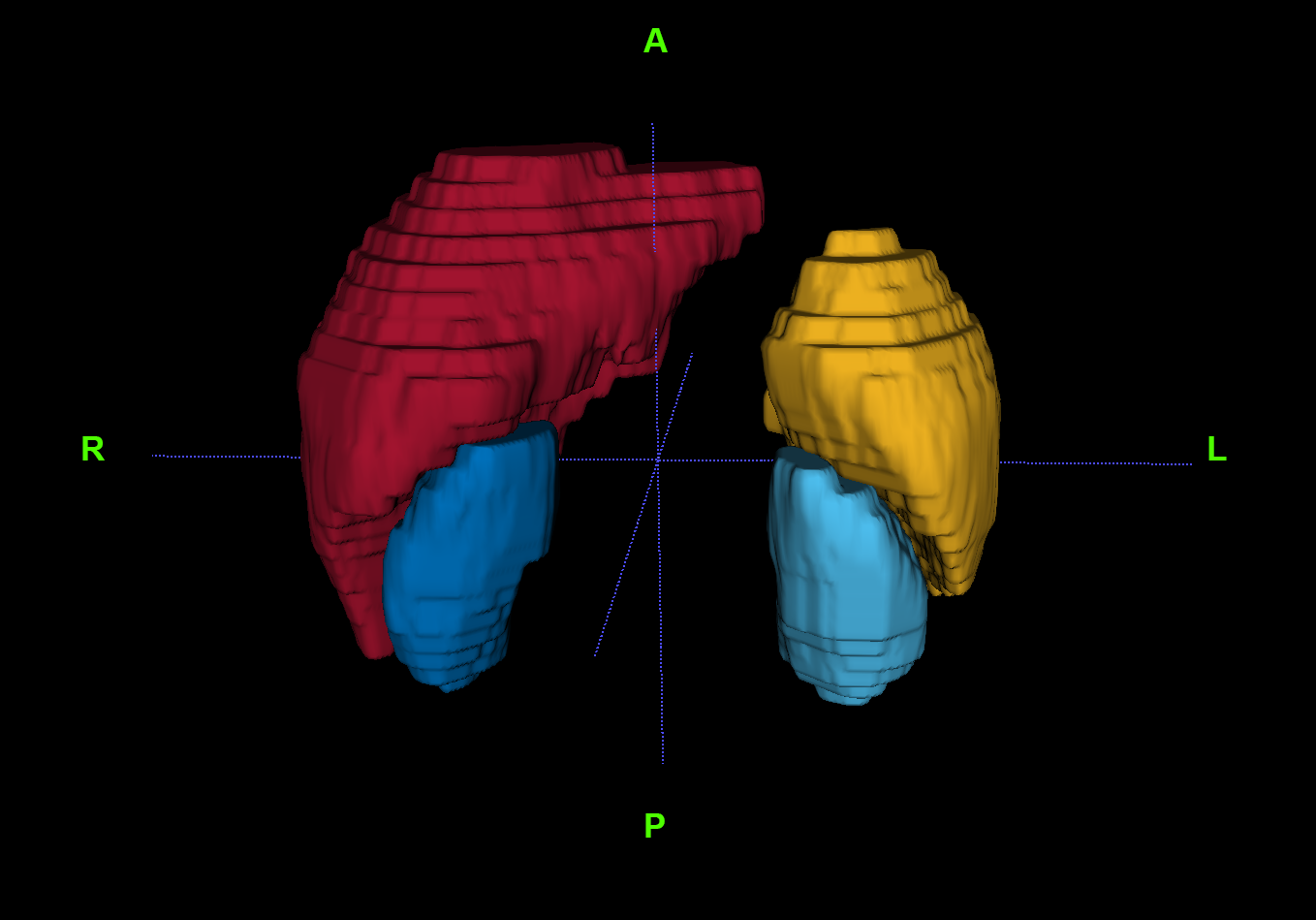}
        \caption{3D visualization of liver (red), right kidney (dark blue), left kidney (light blue) and spleen (yellow) from the CHAOS MR dataset (case 40).}
\label{fig:mr_sample} 
\end{figure}

\begin{table*}[!b]
\def\arraystretch{1.3}
\centering
\caption{Statistics about the CHAOS CT and MRI datasets.}
\label{tab:data_stats}
\begin{tabular}{@{}L{9cm} C{2.5cm} C{2.5cm}}
\toprule 
Specification & CT & MR \\ \midrule 
Number of patients (Train + Test) & 20 + 20 & 20 + 20\\ 
Number of sets (Train + Test) & 20 + 20 & 60 + 60*\\ 
In-plane spatial resolution  & 512 x 512 & 256 x 256  \\ 
Number of axial slices in each examination [min-max] & [78 - 294] & [26 - 50] \\ 
Average axial slice number & 160 & 32x3*\\ 
Total axial slice number & 6407 & 3868x3*\\ 
X spacing (mm/voxel) left-right [min-max] & [0.54 - 0.79] & [0.72 - 2.03]\\ 
Y spacing (mm/voxel) anterior-posterior [min-max] & [0.54 - 0.79] & [0.72 - 2.03]\\ 
Slice thickness (mm) [min-max] & [2.0 - 3.2] & [4.4 - 8.0]\\ 
\bottomrule
\multicolumn{3}{@{}p{14cm}}{\footnotesize{* MRI sets are collected from 3 different pulse sequences. For each patient, T1-DUAL registered \emre{in-phase and oppose-phase} and T2-SPIR MRI data are acquired.}}
\end{tabular}
\end{table*}

\subsubsection{MRI Data Specifications}
The MRI dataset includes two different sequences (T1 and T2) for 40 patients. In total, there are 120 DICOM datasets from T1-DUAL in-phase (40 datasets), oppose-phase (40 datasets), and T2-SPIR (40 datasets). 
\lk{Each of these sets was acquired from routine screening of the abdomen in the clinic.} T1-DUAL \emre{in-phase and oppose-phase} images are registered. Therefore, their \emre{ground truth is} the same. On the other hand, T1 and T2 sequences are not registered. The datasets were acquired on a 1.5T Philips MRI, which produces 12-bit DICOM images. The details of this dataset are given in Tab.\ref{tab:data_stats} and a sample case is illustrated in Fig.\ref{fig:samples}, middle and right, and Fig.\ref{fig:mr_sample}.

\subsection{Aims and Tasks}
The CHAOS challenge has three separate but related aims:

\begin{enumerate}
\item segmentation of the liver from CT scans,
\item segmentation of solid abdominal organs (liver, spleen, kidneys) from MRI sequences.
\item segmentation of organs from mixed (CT-MRI) datasets.
\end{enumerate}

CHAOS provides different \lk{opportunities for} segmentation algorithm design to the participants through five individual tasks:

\textbf{Task 1: Liver Segmentation (CT-MRI)} focuses on using a single system that can segment the liver from both CT and multi-modal MRI (T1-DUAL and T2-SPIR sequences). This corresponds to \quot{cross-modality} learning, which is expected to be used more frequently as the abilities of DL are \emre{improving} \citep{Valindria2018}. 

\textbf{Task 2: Liver Segmentation (CT)} covers a regular segmentation task, which can be considered relatively \emre{easy} due to the inclusion of only healthy livers aligned in the same direction and patient position. On the other hand, the diffusion of contrast agent to parenchyma and the enhancement of the inner vascular tree creates challenging difficulties.

\textbf{Task 3: Liver Segmentation (MRI)} has a similar objective \lk{to} Task 2, but targets multi-modal MRI data randomly collected within \emre{the} routine clinical workflow. The methods are expected to work on both T1-DUAL (in-phase and oppose-phase) as well as T2-SPIR MR sequences.

\textbf{Task 4: Segmentation of abdominal organs (CT-MRI)} is similar to Task 1 with an extension to multiple organ segmentation from MR. In this task, the interesting part is that only the liver is annotated as ground truth in the CT datasets, but the MRI datasets have four annotated abdominal organs. In other words, model input is the same (i.e. 2D slice image or 3D volume), but the number of outputs is different for CT and MR. Such a task is added to the challenge because, in the routine clinical workflow, the aim of acquisition can vary: it can be for multiple organs or a single one. When the scan is performed for a single organ, the remaining organs might not be acquired completely. Thus, a model, which would be used for abdominal organ segmentation in daily workflow, should handle these varying conditions and output types. 

\textbf{Task 5: Segmentation of abdominal organs (MRI)} is the same as Task 3 but extended to four abdominal organs.

Here, it is important to point out that Task 1 can be seen as a union of Tasks 2 and 3. Similarly, Task 4 can be seen as a union of Tasks 1/2 and 5. In that case, the teams might externally divide the datasets into two (i.e. CT and MR originated) and \lk{feed} them separately to the systems they have developed for Tasks 1 and 4. However, the main aim of these tasks is obtaining systems that can handle variations \lk{of} different datasets and better fit to the clinical workflow. Accordingly, for Tasks 1 and 4, a fusion of individual models obtained from different modalities (i.e. two models, one working on CT and the other on MRI) is not valid. In more detail, it is not allowed to combine systems that are specifically set for a single modality and operate completely independently. Instead, novel model designs and better training strategies are expected to handle the challenges associated with these tasks. Alternatively, for Task 4, the fusion of individual modality-specific models can be used if a shared pre-processing block detects modality type and processes the output by different sub-systems. However, this is not valid for Task 1, which specifically aims \lk{at} cross-modality training. Besides, the fusion of individual models for MRI sequences (T1-DUAL and T2-SPIR) is allowed in all MRI-included tasks due to the lower spatial dimension of the MR scans. More details about the tasks are available on the CHAOS challenge website.\footnote{CHAOS description: https://chaos.grand-challenge.org/} \footnote{CHAOS FAQ: https://chaos.grand-challenge.org/News\_and\_FAQ/}

\subsection{Annotations for reference segmentation}
All 2D slices were labeled manually by three different radiology experts who have 10, 12, and 28 years of experience, respectively. The final shapes of the reference segmentations were decided by majority voting. Also, in some extraordinary situations such as when inferior vena cava (IVC) is accepted as a part of the liver, experts have made joint decisions. In CHAOS, voxels that belong to IVC were excluded unless they are not completely inside the liver. Although this handcrafted annotation process has taken a \emre{considerable} amount of time, it was \lk{carried out} to create a consistent and consensus-based ground truth image series. 

\subsection{Challenge Setup and Distribution of the Data}
Both CT and MRI datasets were divided into 20 sets for training and 20 sets for testing. When dividing the sets into training and testing, attention was paid to the fact that the cases in both sets contain similar features (resolution, slice thickness, age of patients) as stratification criteria. We presented \lk{to the CHAOS participants} training data with ground truth labels, \lk{and} test data \lk{containing only the} original images. To provide sufficient data that contains enough variability, the datasets in the training data were selected to represent all the difficulties that are observed on the whole database, such as varying Hounsfield range and non-homogeneous parenchyma texture of \emre{the} liver due to the injection of contrast media in CT images, sudden changes in planar view, and the effect of bias field in MR images.

The images \lk{were} distributed as DICOM files to present the data in its original form. The only modification was removing patient-related information for anonymization. The \emre{ground truth} \lk{was} also presented as image series to match the original format. CHAOS data can be accessed with its DOI number via the zenodo.org webpage under CC-BY-SA 4.0 license \citep{Kavur2019}. One of the important aims of the challenges is to provide data for long-term academic studies. We expect that this data will be used not only for the CHAOS challenge but also for other scientific studies such as cross-modality work or medical image synthesis from different modalities.

\section{Evaluation} 

\subsection{Metrics}
Since the outcomes of medical image segmentation are used for various clinical procedures, using a single metric for 3D segmentation evaluation is not a proper approach to ensure acceptable results for all requirements ~\citep{Maier-Hein2018,Yeghiazaryan2015}. Thus, in the CHAOS challenge, four metrics \lk{were} combined. The metrics \lk{were} chosen among \emre{the most frequently used} \lk{ones} in previous challenges~\citep{Maier-Hein2018}\lk{. Their purpose was} to analyze results in terms of overlapping, volumetric, and spatial differences \lk{between a solution and the ground truth}. Distance measures were transformed to millimeters according to an affine transform matrix which was calculated by attributes (pixel spacing, patient image position, patient image orientation) from DICOM metadata.

Let us assume that $S$ represents \lk{ the set of} voxels in a segmentation result, \lk{and} $G$, \lk{the set of} voxels in the ground truth. The utilized metrics are as follows:
\begin{enumerate}
\item  DICE coefficient (DICE) is calculated as  $2|S\cap G|/(|S|+|G|)$, where  $|.|$ denotes cardinality (the larger, the better).

\item  Relative absolute volume difference (RAVD) compares two volumes in percent. $RAVD = (\rm{abs}(|S|-|G|)/|G|) \times 100$, where `abs' denotes the absolute value (the smaller, the better). 

\item  Average symmetric  surface distance (ASSD) is the average Hausdorff distance between border voxels in $S$ and $G$. The unit of this metric is millimeters (the smaller, the better). 

\item  Maximum symmetric  surface distance (MSSD) is the maximum Hausdorff distance between border voxels in $S$ and $G$. The unit of this metric is millimeters (the smaller, the better).
\end{enumerate}

\subsection{Scoring System}
In the literature, there are mainly two ways of ranking results via multiple metrics. One way is ordering the results by metrics' statistical significance with respect to all results. Another way is converting the metric outputs to the same scale and averaging all ~\citep{Langville2013}. In CHAOS, we adopted the second approach. Values coming from each metric have been transformed to span the interval $[0, 100]$ so that higher values correspond to better segmentation. For this transformation, it was reasonable to apply thresholds in order to cut off unacceptable results and increase the sensitivity of the corresponding metric. We are aware of the fact that decisions on metrics and thresholds have a very critical impact on ranking~\citep{Maier-Hein2018}. Therefore, instead of determining the threshold in an ad-hoc manner, we used intra- and inter-annotator scores obtained from the experts, who created the ground truth. 

The radiologists repeated the annotation process of the five \lk{abdomen scans} for both CT and MRI (i.e. 10 datasets in total) two times to enable intra-annotator variability analysis. These reference masks were used for the calculation of \lk{the} challenge metrics in a pair-wise manner. In Tab.\ref{tab:intra}, all metrics were calculated among repeatedly labeled patient sets for each annotator individually. The average is used to observe the intra-annotator variability. In Tab.\ref{tab:inter}, we run all metrics among patient sets that were annotated by different experts. Their averages are given in the table.

\begin{table}[htb]
\centering
\def\arraystretch{1.3}
\caption{Metrics between two ground truth masks generated by the same annotators (A1, A2, A3) over 5 CT and 5 MRI sets in order to observe intra-annotator variability.}
\label{tab:intra}
\resizebox{0.49\textwidth}{!}{%
\begin{tabular}{lccc}
\toprule 
          & A1                & A2                & A3                \\ \midrule
CT - DICE      & 0.979 $\pm$ 0.013 & 0.982 $\pm$ 0.011 & 0.971 $\pm$ 0.019 \\
CT - RAVD (\%) & 0.358 $\pm$ 0.133 & 0.339 $\pm$ 0.105 & 0.344 $\pm$ 0.112 \\
CT - ASSD (mm) & 0.289 $\pm$ 0.108 & 0.257 $\pm$ 0.102 & 0.243 $\pm$ 0.113 \\
CT - MSSD (mm) & 5.783 $\pm$ 2.154 & 5.756 $\pm$ 2.098 & 5.357 $\pm$ 2.127 \\ \midrule
MRI - DICE      & 0.968 $\pm$ 0.035 & 0.976 $\pm$ 0.076 & 0.969 $\pm$ 0.192 \\
MRI - RAVD (\%) & 0.438 $\pm$ 0.191 & 0.408 $\pm$ 0.312 & 0.472 $\pm$ 0.394 \\
MRI - ASSD (mm) & 0.464 $\pm$ 0.155 & 0.423 $\pm$ 0.440 & 0.412 $\pm$ 0.421 \\
MRI - MSSD (mm) & 6.113 $\pm$ 2.961 & 6.147 $\pm$ 5.903 & 6.057 $\pm$ 5.918 \\
\bottomrule
\end{tabular}%
}
\end{table}
\begin{table}
\centering
\def\arraystretch{1.3}
\caption{Metrics between two ground truth masks generated between annotator pairs (A1 and A2, A1 and A3, A2 and A3) over all patient sets in order to observe inter-annotator variability.}
\label{tab:inter}
\resizebox{0.49\textwidth}{!}{%
\begin{tabular}{lcccc}
\toprule 
     & A1 and A2 & A1 and A3 & A2 and A3 \\ \midrule
DICE & 0.952 $\pm$ 0.098    & 0.949 $\pm$ 0.092    & 0.961 $\pm$ 0.091    \\
RAVD (\%) & 1.525 $\pm$  0.125    & 1.569 $\pm$ 0.118    & 1.465 $\pm$ 0.119    \\
ASSD (mm) & 1.622 $\pm$  0.961   & 1.564 $\pm$ 0.989    & 1.492 $\pm$ 0.9574    \\
MSSD (mm) & 9.174 $\pm$  4.487   & 9.028 $\pm$ 0.428    & 8.877 $\pm$ 0.421  \\
\bottomrule
\end{tabular}%
}
\end{table}

The differences between intra- and inter-annotator variability show the amount of performance change when the same annotation process is repeated by the same expert at a different time or by another expert, respectively. According to Tables \ref{tab:intra} and \ref{tab:inter}, the amount of change for inter- and intra-annotator cases depends on the chosen metric and modality type. Regarding the modality type, the small spacing and inter-slice distance of CT allow a narrow range compared to MRI. 
Regarding the dependence on the chosen metric, for example, DICE, change between intra- and inter-annotator variability is relatively small. On the other hand, the changes in other metrics are observed to be higher.  Based on these analyses, the thresholds are determined by discussions among physicians and computer scientists. As a result, the thresholds were determined as given in Tab.\ref{tab:metric-table}. (The effects of thresholds on ranking stability and robustness is discussed in Section 6.7)


\begin{table}[htb]
\centering
\def\arraystretch{1.3}
\caption{Summary of the metrics and thresholds. $\Delta$ represents longest possible distance in the 3D volume.}
\label{tab:metric-table}
\begin{tabular}{c c c c }
\toprule 
Metric name & Best value & Worst value & Threshold\\ \midrule 
DICE        & 1                  & 0                   & DICE \textgreater 0.8 \\ 
RAVD        & 0\%                & $\infty$               & RAVD \textless 5\%    \\ 
ASSD        & 0 mm               & $\Delta$                & ASSD \textless 15 mm  \\ 
MSSD        & 0 mm               & $\Delta$                 & MSSD \textless 60 mm  \\ 
\bottomrule
\end{tabular}
\end{table}

The metric values outside the threshold range get zero points. The values within the range  are mapped to the interval $[0, 100]$. Then, the scores of each case in the test data \emre{are} calculated as the mean of the four scores. The missing cases (sets that do not have segmentation results) get zero points and these points are included in the final score calculation. The average of the scores across all test cases determines the overall score of the team for the specified task. The code for all metrics (in MATLAB, Python, and Julia) is available at \url{https://github.com/emrekavur/CHAOS-evaluation}. Also, more details about the metrics, the CHAOS scoring system, and a mini-experiment that compares sensitivities of different metrics to distorted segmentations are provided on the same website.

\begin{table*}[!]
\def\arraystretch{1.3}
\centering
\caption{CHAOS challenge submission statistics for on-site and online sessions (between 11 April 2019 - 1 October 2020).}
\label{tab:submissions}
\begin{tabular}{p{9.0cm}cccccc}
\toprule 
Submission numbers & Task~1 & Task~2 & Task~3 & Task~4 & Task~5\\ \midrule 
On-site & 5 & 14 & 7 & 4 & 5 \\ 
Online & 30 & 379 & 178 & 29 & 187 \\ 
Maximum number of submissions by one team (On-site) & 1 & 5 & 1 & 1 & 1 \\ 
Maximum number of submissions by one team (Online) & 3 & 12 & 10 & 5 & 9 \\ 
\bottomrule
\end{tabular}
\end{table*}

\section{Participating Methods} 
In this section, we present the majority of the results from the conference participants and the best two of the post-conference results collected among the online submissions. To be specific, Metu\_MMLab and nnU-Net results belong to online submissions while others are from the conference session. Statistics about submission numbers are presented in Tab.\ref{tab:submissions}. Each method is assigned a unique color code as shown in the figures and tables. The majority of the applied methods (i.e. all except IITKGP-KLIV) used variations of U-Net,  which is a Convolutional Neural Networks (CNN) approach that was first proposed by \cite{ronneberger2015u} for segmentation on biomedical images. This seems to be a typical situation as the corresponding architecture dominates most of the recent DL based segmentation studies even in the presence of limited annotated data which is a typical scenario for biomedical image applications. Among all \lk{studies}, two rely on ensembles (i.e. MedianCHAOS and nnU-Net), which uses multiple models and combine their results.

The following paragraphs, \lk{summarize the} participants' methods. Brief comparisons of them in terms of methodological details and training strategy are given in Tab. \ref{tab:allmethods}. Also, pre-, post-processing and data augmentation strategies are provided in Tab. \ref{tab:preprocessing}.

\teamcolor{OvGUMEMoRIAL}~\textbf{OvGUMEMoRIAL:} A modified version of Attention U-Net proposed in \citep{abraham2019novel} is used. Differently from the original UNet architecture \citep{ronneberger2015u}, in Attention U-Net \citep{abraham2019novel}, soft attention gates are used, a multiscaled input image pyramid is employed for better feature representation, and Tversky loss is computed for the four different scaled levels. The modification adopted by the OvGUMEMoRIAL team is that they employed parametric ReLU activation function instead of ReLU, where an extra parameter, i.e., coefficient of leakage, is learned during training. The ADAM optimizer is used; training is accomplished by 120 epochs with a batch size of 256.

\teamcolor{ISDUE}~\textbf{ISDUE:} The proposed architecture is constructed by three main modules, namely 1) a convolutional autoencoder network which is composed of the \textit{prior encoder} $f_{enc_{p}}$, and decoder $g_{dec}$; 2) a segmentation hourglass network which is composed of the \textit{imitating encoder} $f_{enc_{i}}$, and decoder $g_{dec}$; 3) U-Net module, i.e. $h_{unet}$, which is used to enhance the decoder $g_{dec}$ by guiding the decoding process for better localization capabilities. The segmentation networks, i.e. U-Net module and hourglass network module, are optimized separately using the DICE loss and regularized by $\mathcal{L}_{sc}$ with a regularization weight of $0.001$. The autoencoder is optimized separately using DICE loss. The ADAM optimizer is used with initial learning rate of 0.001, batch size of 1 is used and 2400 iterations are performed to train each model. Data augmentation is performed by applying random translation and rotation operations during training. 

\teamcolor{lachinov}~\textbf{Lachinov:} The proposed model is based on the 3D U-Net architecture, with skip connections between contracting and expanding paths and exponentially growing number of channels across consecutive spatial resolution levels. The encoding path is constructed by a residual network which provides efficient training. Group normalization \citep{wu2018group} is adopted instead of the batch normalization \citep{ioffe2015batch}, by assigning the number of groups to 4. Data augmentation is applied by performing random mirroring of the first two axes of the cropped regions which is followed by random 90 degrees rotation along the last axis and intensity shift with contrast augmentations.

\teamcolor{IITKGP-KLIV}~\textbf{IITKGP-KLIV:} In order to accomplish multi-modality segmentation using a single framework, a multi-task adversarial learning strategy is employed to train a base segmentation network SUMNet \citep{nandamuri2019sumnet}  with batch normalization. To perform adversarial learning, two auxiliary classifiers, namely C1 and C2, and a discriminator network, i.e. D, are used. C1 is trained by the input from the encoder part of SUMNet which provides modality-specific features. A C2 classifier is used to predict the class labels for the selected segmentation maps. The segmentation network and classifier C2 are trained using cross-entropy loss while the discriminator D and auxiliary classifier C1 are trained by binary cross-entropy loss. The ADAM optimizer is used for optimization. The input data to the network is the combination of all four modalities, i.e. CT, MRI T1-DUAL in-phase, and oppose-phase as well as MRI T2-SPIR.    

\teamcolor{METU_MMLAB}~\textbf{METU\_MMLAB:} This model is also designed as a variation of U-Net. In addition, a Conditional Adversarial Network (CAN) is introduced in the proposed model. Batch Normalization is performed before convolution. In this way, vanishing gradients are prevented and selectivity is increased. Moreover, parametric ReLU is employed to preserve the negative values using a trainable leakage parameter. In order to improve the performance around the edges, a CAN is employed during training (not as a post-process operation). This introduces a new loss function to the system which regularizes the parameters for sharper edge responses. Normalization of each CT image is performed for pre-processing and 3D connected component analysis is utilized for post-processing.

\teamcolor{PKDIA}~\textbf{PKDIA:} The team proposed an approach based on conditional generative adversarial networks where the generator is constructed by cascaded partially pre-trained encoder-decoder networks \citep{conze2020abdominal} extending the standard U-Net \citep{ronneberger2015u} architecture. More specifically, first, the standard U-Net encoder part is exchanged for a deeper network, i.e. VGG-19 
by omitting the top layers. Differently from the standard U-Net \citep{ronneberger2015u}, 1) 64 channels (32 channels for standard U-Net) are generated by the first convolutional layer; 2) after each max-pooling operation, the number of channels doubles until it reaches 512 (256 for standard U-Net); 3) second max-pooling operation is followed by 4 consecutive convolutional layers instead of 2. For training, the ADAM optimizer with a learning rate of $10^{-5}$ is used. The fuzzy DICE score is employed as the loss function.

\teamcolor{MedianCHAOS1}~\teamcolor{MedianCHAOS2}~\teamcolor{MedianCHAOS3}~\teamcolor{MedianCHAOS4}~\teamcolor{MedianCHAOS5}~\teamcolor{MedianCHAOS6}~\textbf{MedianCHAOS:} Averaged ensemble of five different networks is used. The first one is \textit{the DualTail-Net architecture} that is composed of an encoder, central block, and 2 dependent decoders. While performing downsampling by max-pooling operation, the max-pooling indices are saved for each feature map to be used during the upsampling operation. The decoder is composed of two branches: one that consists of four blocks and starts from the central block of the U-net architecture, and another one that consists of 3 blocks and starts from the last encoder block. These two branches are processed in parallel where the corresponding feature maps are concatenated after each upsampling operation. The decoder is followed by a $1\times1$ convolution and sigmoid activation function which provides a binary segmentation map at the output.

The other four networks are U-Net architecture variants, i.e. TernausNet (U-Net with VGG11 backbone \citep{iglovikov2018ternausnet}), LinkNet34  \citep{shvets2018automatic}, and two networks with ResNet-50 and SE-Resnet50. The latter two were both pretrained on ImageNet encoders and decoders and consist of convolution, ReLU, and transposed convolutions with stride 2. The two best final submissions were the averaged ensembles of predictions obtained by these five networks. The training process for each network was performed with the ADAM optimizer. DualTail-Net and LinkNet34 were trained with soft DICE loss and the other three networks were trained with the combined loss: 0.5*soft DICE + 0.5*BCE (binary cross-entropy). No additional post-processing was performed.

\begin{table*}[p]
\caption{\label{tab:allmethods} Brief comparison of participating methods}
\centering
\scriptsize
\def\arraystretch{1.3}
\begin{tabular}{>{\raggedright}p{2.4cm}p{8.2cm}p{6.4cm}}
\toprule 
\multicolumn{1}{c}{\textbf{Team}} & \multicolumn{1}{c}{\textbf{Details of the method}} &\multicolumn{1}{c}{\textbf{Training strategy}}\\ \midrule
\teamcolor{OvGUMEMoRIAL} \textbf{OvGUMEMoRIAL} \tiny{(P.~Ernst, S.~Chatterjee, O.~Speck, A.~Nürnberger)} &  
    \textbullet Modified Attention 2D U-Net \citep{abraham2019novel}, employing {soft attention gates} and multiscaled input image pyramid for better feature representation is used.\newline     
    
    \textbullet Parametric ReLU activation is used instead of ReLU, where an extra parameter, i.e. coefficient of leakage, is learned during training. 
 & 
     \textbullet Tversky loss is computed for the four different scaled levels. \newline
     
     \textbullet The ADAM optimizer is used, training is accomplished by 120 epochs with a batch size of 256.
  \\ \midrule
\teamcolor{ISDUE} \textbf{ISDUE} \newline \tiny{(D.~D.~Pham, G.~Dovletov, J.~Pauli)} & 
    \textbullet  The proposed architecture consists of three main modules:\newline      

    i. Autoencoder net composed of a \textit{prior encoder} $f_{enc_{p}}$, and decoder $g_{dec}$; \newline      
    ii. Hourglass net composed of an \textit{imitating encoder} $f_{enc_{i}}$, and decoder $g_{dec}$; \newline     
    iii.  2D U-Net module, i.e. $h_{unet}$, which is used to enhance the decoder $g_{dec}$ by guiding the decoding process for better localization capabilities.

& 
    \textbullet The segmentation networks are optimized separately using the DICE-loss and regularized by $\mathcal{L}_{sc}$ with weight of $\lambda=0.001$. \newline
    
    \textbullet The autoencoder is optimized separately using DICE loss.\newline 
    
    \textbullet The ADAM optimizer with an initial learning rate of 0.001, and 2400 iterations are performed to train each model.
\\  \midrule
\teamcolor{lachinov} \textbf{Lachinov} \newline \tiny{(D.~Lachinov)} & 

    \textbullet 3D U-Net, with skip connections between contracting/expanding paths and exponentially growing number of channels across the consecutive resolution levels \citep{Lachinov19}.\newline  
    
 \textbullet The encoding path is constructed by a residual network for efficient training.\newline     
 \textbullet Group normalization \citep{wu2018group} is adopted instead of batch \citep{ioffe2015batch} (\# of groups = 4).\newline  
 
 \textbullet Pixel shuffle is used as an upsampling operator  
    & 
    \textbullet  The network was trained with ADAM optimizer with learning rate 0.001 and decaying with a rate of 0.1 at the 7th and 9th epoch.\newline   
    
    \textbullet The network is trained with batch size 6 for 10 epochs. Each epoch has 3200 iterations in it.\newline     
    
     \textbullet The loss function employed is DICE loss.
\\ \midrule
\teamcolor{IITKGP-KLIV} \textbf{IITKGP-KLIV} \tiny{(R.~Sathish, R.~Rajan, D.~Sheet)} & 
    \textbullet To achieve multi-modality segmentation using a single framework, a multi-task adversarial learning strategy is employed to train a base segmentation network 2D SUMNet \citep{nandamuri2019sumnet} with batch normalization.\newline
    
    \textbullet  Adversarial learning is performed by two auxiliary classifiers, namely C1 and C2, and a discriminator network D.    
& 

    \textbullet The segmentation network and C2 are trained using cross-entropy loss while the discriminator D and auxiliary classifier C1 are trained by binary cross-entropy loss. \newline     
    
    \textbullet The ADAM optimizer. Input is the combination of all four modalities, i.e. CT, MRI T1 DUAL In-phase and Oppose-phase MRI T2 SPIR.\\ \midrule
\teamcolor{METU_MMLAB} \textbf{METU\_MMLAB} \tiny{(S.~Özkan, B.~Baydar, G.~B.~Akar)} & 
    \textbullet A 2D U-Net variation and a Conditional Adversarial Network (CAN) is introduced.\newline   
    
  \textbullet  Batch Normalization is performed before convolution to prevent vanishing gradients and increase selectivity.\newline 
  
 \textbullet Parametric ReLU to preserve negative values using a trainable leakage parameter.   

& 

    \textbullet To improve the performance around the edges, a CAN is employed during training (not as a post-process operation).\newline  
    
    \textbullet This introduces a new loss function to the system which regularizes the parameters for sharper edge responses. 
 \\ \midrule
 
\teamcolor{PKDIA} \textbf{PKDIA} \tiny{(P.-H.~Conze)} & 
    \textbullet An approach based on Conditional Generative Adversarial Networks (cGANs) is proposed: the generator is built by cascaded pre-trained encoder-decoder (ED) networks \citep{conze2020abdominal} extending the standard 2D U-Net (sU-Net) \citep{ronneberger2015u} (VGG19, following \citep{conze2019healthy}), with 64 channels (instead of 32 for sU-Net) generated by the first convolutional layer. \newline
    
    \textbullet After each max-pooling, the channel number doubles until 512 (256 for sU-Net). Max-pooling followed by 4 consecutive conv. layers instead of 2. The auto-context paradigm is adopted by cascading two EDs \citep{Yan2019}: the output of the first is used as features for the second.
 & 
    \textbullet The ADAM optimizer with a learning rate of $10^{-5}$ is used.\newline   
    
    \textbullet The fuzzy DICE score is employed as a loss function. \newline  
    
    \textbullet The batch size was set to 3 for CT and 5 for MR scans.
 \\ \midrule
\teamcolor{MedianCHAOS1}\teamcolor{MedianCHAOS2}\teamcolor{MedianCHAOS3}\teamcolor{MedianCHAOS4}\teamcolor{MedianCHAOS5}\teamcolor{MedianCHAOS6} \textbf{MedianCHAOS} \tiny{(V.~Groza)} & 
    \textbullet Averaged ensemble of five different networks is used. The first one is \textit{DualTail-Net} that is composed of an encoder, central block, and 2 dependent decoders.\newline     
    
    \textbullet The other four networks are U-Net variants, i.e. TernausNet (2D U-Net with VGG11 backbone \citep{iglovikov2018ternausnet}), LinkNet34  \citep{shvets2018automatic}, and two with ResNet-50 and SE-Resnet50. 
      & 
        \textbullet The training for each network was performed with the ADAM.\newline   
        
        \textbullet  DualTail-Net and LinkNet34 were trained with soft DICE loss and the other three networks were trained with the combined loss: 0.5*soft DICE + 0.5*BCE (binary cross-entropy). 
\\ \midrule
\teamcolor{mountain} \textbf{Mountain} \tiny{(Shuo~Han)} &

    \textbullet 3D network adopting the U-Net variant in \citep{han2019cerebellum} is used. It differs from U-Net in \citep{ronneberger2015u}, by adopting:
        i. A pre-activation residual block in each scale level at the encoder,
        ii. Convolutions with stride 2 to reduce the spatial size,
        iii. Instance normalization \citep{ulyanov2017improved}.\newline  
        
    \textbullet Two nets, i.e. NET1 and NET2, adopting \citep{han2019cerebellum} with different channels and levels. NET1 locates organ and outputs a mask for NET2 performing finer segmentation.
 & 
    \textbullet The ADAM optimizer is used with the initial learning rate $ = 1 \times 10 ^ {-3}$, $\beta_1 = 0.9 $, $\beta_2 = 0.999$, and $ \epsilon = 1 \times 10 ^ {-8}$. \newline
    
    \textbullet DICE coefficient was used as the loss function. The batch size was set to 1.
\\ \midrule
\teamcolor{CIR-MPerkonigg} \textbf{CIRMPerkonigg} \tiny{(M.~Perkonigg)} & 
    \textbullet For joint training with all modalities, the IVD-Net \citep{dolz2018ivd} (which is an extension of 2D U-Net \cite{ronneberger2015u}) is used with a number of modifications: \newline    
    i. dense connections between encoder path of IVD-Net are not used since no improvement is achieved\newline    
    ii. training images are split.\newline 
    
    \textbullet Moreover, residual convolutional blocks \citep{he2016deep} are used.
& 
    \textbullet Modality Dropout \citep{li2016modout} is used as the regularization technique 
    to decrease over-fitting on certain modalities.\newline     
    
    \textbullet Training is done by using the ADAM optimizer with a learning rate of 0.001 for 75 epochs. 
 \\ \midrule
\teamcolor{nnU-Net} \textbf{nnU-Net} \newline \tiny{(F.~Isensee, K.~H.~Maier-Hein)} &
    \textbullet An internal variant of nnU-Net \citep{Isensee2019}, which is the winner of Medical Segmentation Decathlon (MSD) in 2018 \citep{simpson2019large}, is used. \newline
    
    \textbullet The ensemble of five 3D U-Nets (``3d\_fullres'' configuration), which originate from cross-validation on the training cases. Ensemble of T1 in-phase and oppose-phase was used.
& 
    \textbullet T1 in and out are treated as separate training examples, resulting in a total of 60 training examples for the tasks. \newline 
    
    \textbullet Task 3 is a subset of Task 5, so training was done only once and the predictions for Task 3 were generated by isolating the liver. 
 \\ 
\bottomrule
\end{tabular}
\end{table*}

\begin{table*}[!t]
\caption{\label{tab:preprocessing} Pre-processing, post-processing and data augmentation operations together with participated tasks.}
\centering
\footnotesize
\def\arraystretch{1.5}
\begin{tabular}{>{\raggedright}p{2.5cm} >{\raggedright}p{4.2cm} >{\raggedright}p{4.2cm} >{\raggedright}p{4.2cm} l}
\toprule 
Team & Pre-process  & Data augmentation & Post-process & Tasks \\ \midrule
\teamcolor{OvGUMEMoRIAL} OvGUMEMoRIAL & Training with resized images ($128\times128$). Inference: full-sized. & - & Threshold by 0.5 & 1,2,3,4,5  \\ \midrule
\teamcolor{ISDUE} ISDUE & Training with resized images (96,128,128) & Random translate and rotate & Threshold by 0.5. Bicubic interpolation for refinement. & 1,2,3,4,5 \\  \midrule
\teamcolor{lachinov} Lachinov & Resampling $1.4\times1.4\times2$ z-score normalization & Random ROI crop $192\times192\times64$, mirror X-Y, transpose X-Y, Window Level - Window Width & Threshold by 0.5 & 1,2,3 \\  \midrule
\teamcolor{IITKGP-KLIV} IITKGP-KLIV & Training with resized images ($256\times256$), whitening. Additional class for body. & - & Threshold by 0.5 & 1,2,3,4,5\\ \midrule
\teamcolor{METU_MMLAB} METUMMLAB & Min-max normalization for CT & - & Threshold by 0.5. Connected component analysis for selecting/eliminating some of the model outputs. & 1,3,5 \\ \midrule
\teamcolor{PKDIA} PKDIA & Training with resized images:~$256\times256$ MR,~$512\times512$ CT. & Random scale, rotate, shear and shift & Threshold by 0.5. Connected component analysis for selecting/eliminating some of the model outputs. & 1,2,3,4,5\\  \midrule
\teamcolor{MedianCHAOS1}\teamcolor{MedianCHAOS2}\teamcolor{MedianCHAOS3}\teamcolor{MedianCHAOS4}\teamcolor{MedianCHAOS5}\teamcolor{MedianCHAOS6}\newline MedianCHAOS & LUT {[}-240,160{]} HU range, normalization. & - & Threshold by 0.5. & 2 \\ \midrule
\teamcolor{mountain} Mountain & Resampling $1.2\times1.2\times4.8$, zero padding. Training with resized images: $384\times384\times64$. Rigid register MR. & Random rotate, scale, elastic deformation & Threshold by 0.5. Connected component analysis for selecting/eliminating some of the model outputs.  & 3,5 \\ \midrule 
\teamcolor{CIR-MPerkonigg} CIRMPerkonigg & Normalization to zero mean unit variance. & 2D Affine and elastic transforms, histogram shift, flip and adding Gaussian noise. & Threshold by 0.5. & 3 \\  \midrule
\teamcolor{nnU-Net} nnU-Net &  Normalization to zero mean unit variance, Resampling $1.6\times1.6\times5.5$  & Add Gaussian noise / blur, rotate, scale, WL-WW, simulated low resolution, Gamma, mirroring & Threshold by 0.5. & 3,5\\  \bottomrule
\end{tabular}
\end{table*}

\teamcolor{mountain}~\textbf{Mountain:} A 3D network architecture modified from the U-Net in \citep{han2019cerebellum} is used. Differently from U-Net in \citep{ronneberger2015u}, in \citep{han2019cerebellum} a pre-activation residual block in each scale level is used at the encoder part; instead of max pooling, convolutions with stride 2 to reduce the spatial size is employed; and instead of batch normalization, instance normalization \citep{ulyanov2017improved} is used since instance normalization is invariant to linear changes in the intensity of each individual image. Finally, it sums up the outputs of all levels in the decoder as the final output to encourage convergence. Two networks adopting the aforementioned architecture with a different number of channels and levels are used here. The first network, NET1, is used to locate an organ such as the liver.  It outputs a mask of the organ to crop out the region of interest to reduce the spatial size of the input to the second network, NET2. The output of NET2 is used as the final segmentation of this organ. The ADAM optimizer is used with the initial learning rate $ = 1 \times 10 ^ {-3}$, $\beta_1 = 0.9 $, $\beta_2 = 0.999$, and $ \epsilon = 1 \times 10 ^ {-8}$. DICE coefficient was used as the loss function. The batch size was set to 1. Random rotation, scaling, and elastic deformation were used for data augmentation during training.

\teamcolor{CIR-MPerkonigg}~\textbf{CIR\_MPerkonigg:} In order to train the network jointly for all modalities, the IVD-Net architecture of \citep{dolz2018ivd} is employed. It follows the structure of U-Net \citep{ronneberger2015u} with a number of modifications listed as follows: 

1) Dense connections between encoder path of IVD-Net are not used since no improvement is obtained with that scheme, 

2) Not all images are used as input to the network during training. 

Residual convolutional blocks \citep{he2016deep} are used. Data augmentation is performed by accomplishing affine transformations, elastic transformations in 2D, histogram shifting, flipping, and Gaussian noise addition. In addition, Modality Dropout \citep{li2016modout} is used as the regularization technique where modalities are dropped with a certain probability when the training is performed using multiple modalities which helps decrease overfitting on certain modalities. Training is done by using the ADAM optimizer with a learning rate of 0.001 for 75 epochs.

\teamcolor{nnU-Net}~\textbf{nnU-Net:} The nnU-Net team participated in the challenge with an internal variant of nnU-Net \citep{Isensee2019}, which is the winner of Medical Segmentation Decathlon (MSD) in 2018,  \citep{simpson2019large}. They have made submissions for Task 3 and Task 5. These tasks need to process T1-DUAL in-phase and oppose-phase images as well as T2-SPIR images. While the T1-DUAL images are registered and can be used as separate color channel inputs, it was not chosen to do so because this would have required substantial modification to nnU-Net (2 input modalities for T1-DUAL, 1 input modality for T2-SPIR). Instead, T1-DUAL in-phase and oppose-phase were treated as separate training examples, resulting in a total of 60 training examples for the aforementioned tasks. 

No external data was used. Task 3 is a subset of Task 5, so training was done only  once and the predictions for Task 3 were generated by isolating the liver label. The submitted predictions are a result of an ensemble of three 3D U-Nets (``3d\_fullres'' configuration of nnU-Net). The five models originate from cross-validation on the training cases. Furthermore, since only one prediction is accepted for both T1-DUAL image types, an ensemble of the predictions of T1-DUAL in-phase and oppose-phase was used.

\section{Results}

\begin{figure*}[!htb]
    \centering
    \begin{minipage}[t]{1.0\textwidth}
        \centering
        \includegraphics[width=1.0\textwidth]{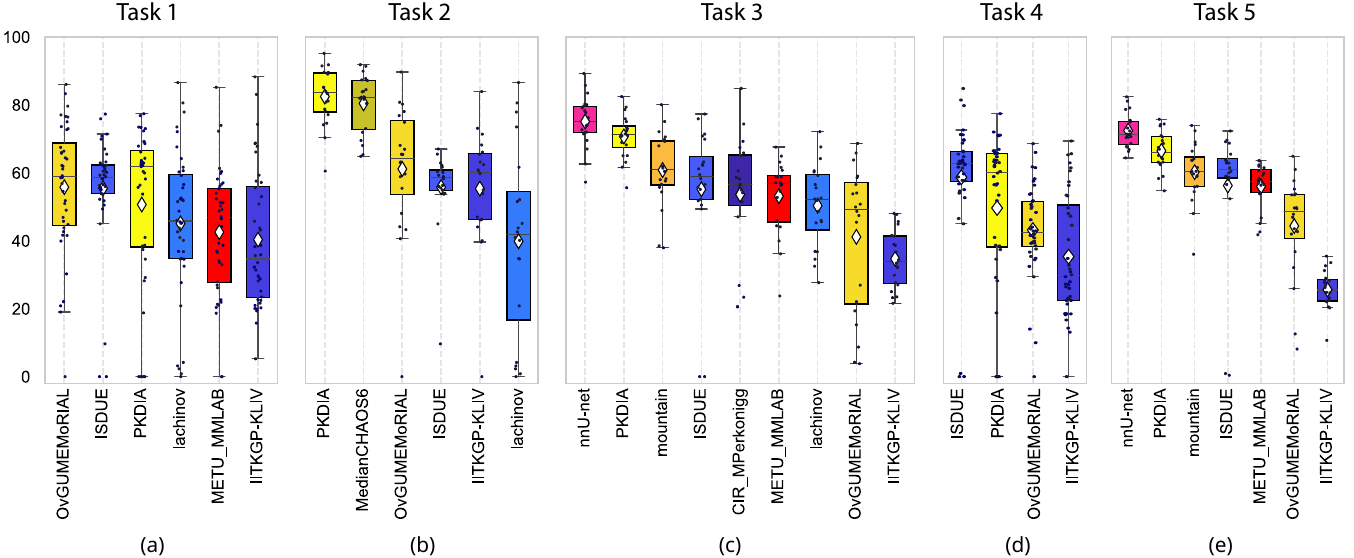}
        \caption{Box plot of the methods' score for (a) Task 1, (b) Task 2, (c) Task 3, (d) Task 4, and (e) Task 5 on test data. White diamonds represent the mean values of the scores. Solid vertical lines inside of the boxes represent medians. Separate dots show scores of each individual case.}
        \label{fig:task_mean}
    \end{minipage}\vspace{20mm}
    \begin{minipage}[t]{1.0\textwidth}
        \centering
        \includegraphics[width=1.0\textwidth]{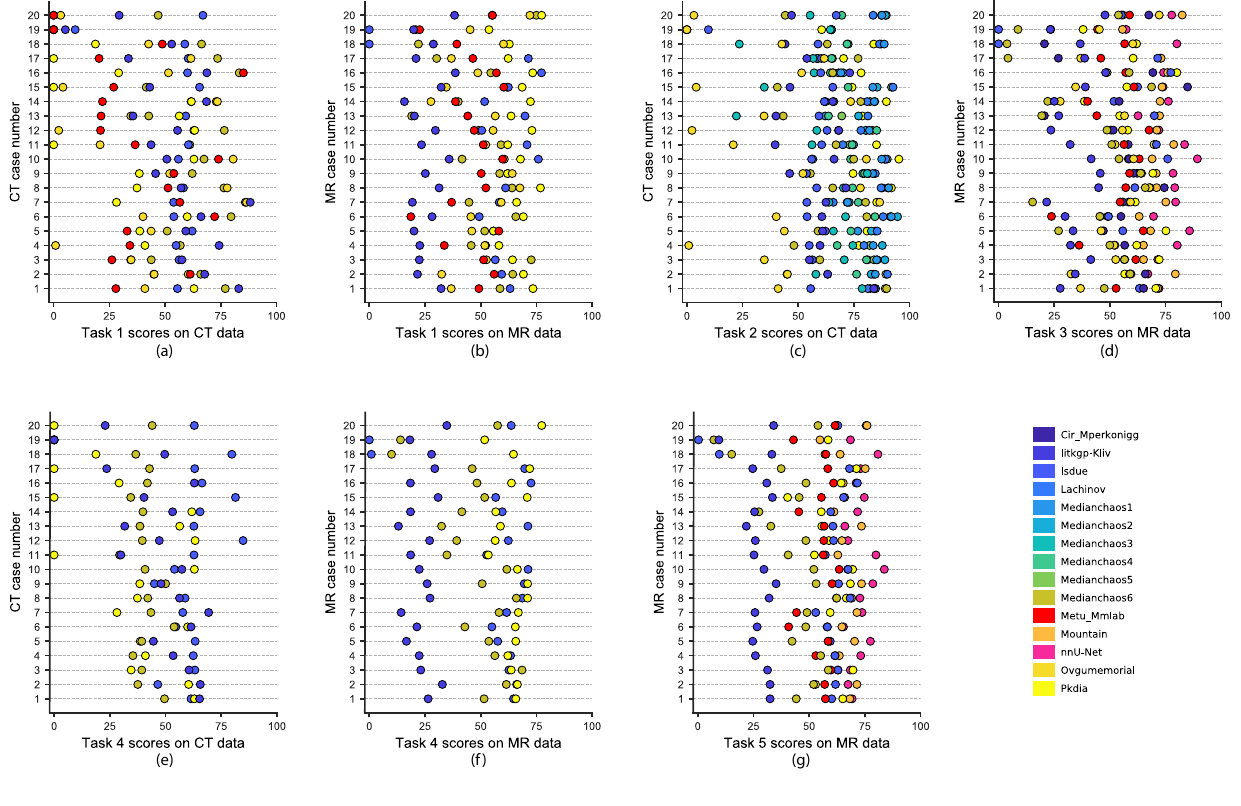}
        \caption{Distribution of the scores for individual cases on test data.}
\label{fig:task_all} 
    \end{minipage}\vspace{5mm}
\end{figure*}

\begin{table*}[!t]
\centering
\caption{Metric values and corresponding scores of submissions. The given values represent the average of all cases and all organs of the related tasks in the test data. The best results are given in bold.}
\label{tab:scores}
\resizebox{\textwidth}{!}{%
\def\arraystretch{1.8}
\begin{tabular}{cllllllllll}
\toprule 
 & Team Name & \textbf{Mean Score} & DICE & DICE Score & RAVD ($\%$) & RAVD Score & ASSD (mm) & ASSD Score & MSSD (mm) & MSSD Score \\ \midrule
\multirow{6}{*}{\rotatebox[origin=c]{90}{\parbox{0.1\textwidth }{\centering \textbf{Task 1}}}}  & \teamcolor{OvGUMEMoRIAL} OvGUMEMoRIAL & \textbf{55.78 $\pm$ 19.20}& 0.88 $\pm$ 0.15 & 83.14 $\pm$ 28.16 & 13.84 $\pm$ 30.26 & \textbf{24.67 $\pm$ 31.15} & 11.86 $\pm$ 65.73 &\textbf{76.31 $\pm$ 21.13} & 57.45 $\pm$ 67.52 & 31.29 $\pm$ 26.01 \\
 & \teamcolor{ISDUE} ISDUE & 55.48 $\pm$ 16.59 & 0.87 $\pm$ 0.16 & 83.75 $\pm$ 25.53 & 12.29 $\pm$ 15.54 & 17.82 $\pm$ 30.53 & 5.17 $\pm$ 8.65 & 75.10 $\pm$ 22.04 & 36.33 $\pm$ 21.97 & 44.83 $\pm$ 21.78 \\
 & \teamcolor{PKDIA} PKDIA & 50.66 $\pm$ 23.95 & 0.85 $\pm$ 0.26 &\textbf{84.15 $\pm$ 28.45} & 6.65 $\pm$ 6.83 & 21.66 $\pm$ 30.35 & 9.77 $\pm$ 23.94 & 75.84 $\pm$ 28.76 & 46.56 $\pm$ 45.02 & \textbf{42.28 $\pm$ 27.05} \\
 & \teamcolor{lachinov} Lachinov & 45.10 $\pm$ 21.91 & 0.87 $\pm$ 0.13 & 77.83 $\pm$ 33.12 & 10.54 $\pm$ 14.36 & 21.59 $\pm$ 32.65 & 7.74 $\pm$ 14.42 & 63.66 $\pm$ 31.32 & 83.06 $\pm$ 74.13 & 24.30 $\pm$ 27.78 \\
 & \teamcolor{METU_MMLAB} METU\_MMLAB & 42.54 $\pm$ 18.79 & 0.86 $\pm$ 0.09 & 75.94 $\pm$ 32.32 & 18.01 $\pm$ 22.63 & 14.12 $\pm$ 25.34 & 8.51 $\pm$ 16.73 & 60.36 $\pm$ 28.40 & 62.61 $\pm$ 51.12 & 24.94 $\pm$ 25.26 \\
 & \teamcolor{IITKGP-KLIV} IITKGP-KLIV \vspace{0.5mm} & 40.34 $\pm$ 20.25 & 0.72 $\pm$ 0.31 & 60.64 $\pm$ 44.95 & 9.87 $\pm$ 16.27 & 24.38 $\pm$ 32.20 & 11.85 $\pm$ 16.87 & 50.48 $\pm$ 37.71 & 95.43 $\pm$ 53.17 & 7.22 $\pm$ 18.68 \\
\midrule 
\multirow{11}{*}{\rotatebox[origin=c]{90}{\parbox{0.1\textwidth }{\centering \textbf{Task 2}}}} & \teamcolor{PKDIA} \textit{PKDIA*} & \textbf{\textit{82.46 $\pm$ 8.47}} & \textit{0.98 $\pm$ 0.00} & \textbf{\textit{97.79 $\pm$ 0.43}} & \textit{1.32 $\pm$ 1.30}2 & \textbf{\textit{73.6 $\pm$ 26.44}} & \textit{0.89 $\pm$ 0.36} & \textbf{\textit{94.06 $\pm$ 2.37}} & \textit{21.89 $\pm$ 13.94} & \textbf{\textit{64.38 $\pm$ 20.17}} \\
 & \teamcolor{MedianCHAOS6} MedianCHAOS6 & 80.45 $\pm$ 8.61 & 0.98 $\pm$ 0.00 & 97.55 $\pm$ 0.42 & 1.54 $\pm$ 1.22 & 69.19 $\pm$ 24.47 & 0.90 $\pm$ 0.24 & 94.02 $\pm$ 1.6 & 23.71 $\pm$ 13.66 & 61.02 $\pm$ 21.06 \\
 & \teamcolor{MedianCHAOS3} MedianCHAOS3 & 80.43 $\pm$ 9.23 & 0.98 $\pm$ 0.00 & 97.59 $\pm$ 0.44 & 1.41 $\pm$ 1.23 & 71.78 $\pm$ 24.65 & 0.9 $\pm$ 0.27 & 94.02 $\pm$ 1.79 & 27.35 $\pm$ 21.28 & 58.33 $\pm$ 21.74 \\
 & \teamcolor{MedianCHAOS1} MedianCHAOS1 & 79.91 $\pm$ 9.76 & 0.97 $\pm$ 0.1 & 97.49 $\pm$ 0.51 & 1.68 $\pm$ 1.45 & 66.8 $\pm$ 28.03 & 0.94 $\pm$ 0.29 & 93.75 $\pm$ 1.91 & 23.04 $\pm$ 10 & 61.6 $\pm$ 16.67 \\
 & \teamcolor{MedianCHAOS2} MedianCHAOS2 & 79.78 $\pm$ 9.68 & 0.97 $\pm$ 0.00 & 97.49 $\pm$ 0.47 & 1.5 $\pm$ 1.2 & 69.99 $\pm$ 23.96 & 0.99 $\pm$ 0.37 & 93.39 $\pm$ 2.48 & 27.96 $\pm$ 23.02 & 58.23 $\pm$ 20.27 \\
 & \teamcolor{MedianCHAOS5} MedianCHAOS5 & 73.39 $\pm$ 6.96 & 0.97 $\pm$ 0.01 & 97.32 $\pm$ 0.41 & 1.43 $\pm$ 1.12 & 71.44 $\pm$ 22.43 & 1.13 $\pm$ 0.43 & 92.47 $\pm$ 2.87 & 60.26 $\pm$ 50.11 & 32.34 $\pm$ 26.67 \\
 & \teamcolor{OvGUMEMoRIAL} OvGUMEMoRIAL & 61.13 $\pm$ 19.72 & 0.90 $\pm$ 0.21 & 90.18 $\pm$ 21.25 & 9x$10^3$ $\pm$ 4x$10^3$ & 44.35 $\pm$ 35.63 & 4.89 $\pm$ 12.05 & 81.03 $\pm$ 20.46 & 55.99 $\pm$ 38.47 & 28.96 $\pm$ 26.73 \\
 & \teamcolor{MedianCHAOS4} MedianCHAOS4 & 59.05 $\pm$ 16 & 0.96 $\pm$ 0.02 & 96.19 $\pm$ 1.97 & 3.39 $\pm$ 3.9 & 50.38 $\pm$ 33.2 & 3.88 $\pm$ 5.76 & 77.4 $\pm$ 28.9 & 91.97 $\pm$ 57.61 & 12.23 $\pm$ 19.17 \\
 & \teamcolor{ISDUE} ISDUE & 55.79 $\pm$ 11.91 & 0.91 $\pm$ 0.04 & 87.08 $\pm$ 20.6 & 13.27 $\pm$ 7.61 & 4.16 $\pm$ 12.93 & 3.25 $\pm$ 1.64 & 78.30 $\pm$ 10.96 & 27.99 $\pm$ 9.99 & 53.60 $\pm$ 15.76 \\
 & \teamcolor{IITKGP-KLIV} IITKGP-KLIV & 55.35 $\pm$ 17.58 & 0.92 $\pm$ 0.22 & 91.51 $\pm$ 21.54 & 8.36 $\pm$ 21.62 & 30.41 $\pm$ 27.12 & 27.55 $\pm$ 114.04 & 81.97 $\pm$ 21.88 & 102.37 $\pm$ 110.9 & 17.50 $\pm$ 21.79 \\
 & \teamcolor{lachinov} Lachinov & 39.86 $\pm$ 27.90 & 0.83 $\pm$ 0.20 & 68.00 $\pm$ 40.45 & 13.91 $\pm$ 20.4 & 22.67 $\pm$ 33.54 & 11.47 $\pm$ 22.34 & 53.28 $\pm$ 33.71 & 93.70 $\pm$ 79.40 & 15.47 $\pm$ 24.15 \\
\midrule 
\multirow{9}{*}{\rotatebox[origin=c]{90}{\parbox{0.1\textwidth }{\centering \textbf{Task 3}}}} & \teamcolor{nnU-Net} nnU-Net & \textbf{75.10 $\pm$ 7.61} & 0.95 $\pm$ 0.01 & \textbf{95.42 $\pm$ 1.32} & 2.85 $\pm$ 1.55 & \textbf{47.92 $\pm$ 25.36} & 1.32 $\pm$ 0.83 & \textbf{91.19 $\pm$ 5.55} & 20.85 $\pm$ 10.63 & \textbf{65.87 $\pm$ 15.73} \\
 & \teamcolor{PKDIA} PKDIA & 70.71 $\pm$ 6.40 & 0.94 $\pm$ 0.01 & 94.47 $\pm$ 1.38 & 3.53 $\pm$ 2.14 & 41.8 $\pm$ 24.85 & 1.56 $\pm$ 0.68 & 89.58 $\pm$ 4.54 & 26.06 $\pm$ 8.20 & 56.99 $\pm$ 12.73 \\
 & \teamcolor{mountain} Mountain & 60.82 $\pm$ 10.94 & 0.92 $\pm$ 0.02 & 91.89 $\pm$ 1.99 & 5.49 $\pm$ 2.77 & 25.97 $\pm$ 27.95 & 2.77 $\pm$ 1.32 & 81.55 $\pm$ 8.82 & 35.21 $\pm$ 14.81 & 43.88 $\pm$ 17.60 \\
 & \teamcolor{ISDUE} ISDUE & 55.17 $\pm$ 20.57 & 0.85 $\pm$ 0.19 & 82.08 $\pm$ 28.11 & 11.8 $\pm$ 15.69 & 24.65 $\pm$ 27.58 & 6.13 $\pm$ 10.49 & 73.50 $\pm$ 25.91 & 40.50 $\pm$ 24.45 & 40.45 $\pm$ 20.90 \\
 & \teamcolor{CIR-MPerkonigg} CIR\_MPerkonigg & 53.60 $\pm$ 17.92 & 0.91 $\pm$ 0.07 & 84.35 $\pm$ 19.83 & 10.69 $\pm$ 20.44 & 31.38 $\pm$ 25.51 & 3.52 $\pm$ 3.05 & 77.42 $\pm$ 18.06 & 82.16 $\pm$ 50 & 21.27 $\pm$ 23.61 \\
 & \teamcolor{METU_MMLAB} METU\_MMLAB & 53.15 $\pm$ 10.92 & 0.89 $\pm$ 0.03 & 81.06 $\pm$ 18.76 & 12.64 $\pm$ 6.74 & 10.94 $\pm$ 15.27 & 3.48 $\pm$ 1.97 & 77.03 $\pm$ 12.37 & 35.74 $\pm$ 14.98 & 43.57 $\pm$ 17.88 \\
 & \teamcolor{lachinov} Lachinov & 50.34 $\pm$ 12.22 & 0.90 $\pm$ 0.05 & 82.74 $\pm$ 18.74 & 8.85 $\pm$ 6.15 & 21.04 $\pm$ 21.51 & 5.87 $\pm$ 5.07 & 68.85 $\pm$ 19.21 & 77.74 $\pm$ 43.7 & 28.72 $\pm$ 15.36 \\
 & \teamcolor{OvGUMEMoRIAL} OvGUMEMoRIAL & 41.15 $\pm$ 21.61 & 0.81 $\pm$ 0.15 & 64.94 $\pm$ 37.25 & 49.89 $\pm$ 71.57 & 10.12 $\pm$ 14.66 & 5.78 $\pm$ 4.59 & 64.54 $\pm$ 24.43 & 54.47 $\pm$ 24.16 & 25.01 $\pm$ 20.13 \\
 & \teamcolor{IITKGP-KLIV} IITKGP-KLIV \vspace{0.5mm}& 34.69 $\pm$ 8.49 & 0.63 $\pm$ 0.07 & 46.45 $\pm$ 1.44 & 6.09 $\pm$ 6.05 & 43.89 $\pm$ 27.02 & 13.11 $\pm$ 3.65 & 40.66 $\pm$ 9.35 & 85.24 $\pm$ 23.37 & 7.77 $\pm$ 12.81 \\
\midrule 
\multirow{4}{*}{\rotatebox[origin=c]{90}{\parbox{0.09\textwidth }{\centering \textbf{Task 4}}}} & \teamcolor{ISDUE} ISDUE & \textbf{58.69 $\pm$ 18.65} & 0.85 $\pm$ 0.21 & 81.36 $\pm$ 28.89 & 14.04 $\pm$ 18.36 & 14.08 $\pm$ 27.3 & 9.81 $\pm$ 51.65 & 78.87 $\pm$ 25.82 & 37.12 $\pm$ 60.17 & 55.95 $\pm$ 28.05 \\
 & \teamcolor{PKDIA} PKDIA & 49.63 $\pm$ 23.25 & 0.88 $\pm$ 0.21 & \textbf{85.46 $\pm$ 25.52} & 8.43 $\pm$ 7.77 & \textbf{18.97 $\pm$ 29.67} & 6.37 $\pm$ 18.96 & \textbf{82.09 $\pm$ 23.96} & 33.17 $\pm$ 38.93 & \textbf{56.64 $\pm$ 29.11} \\
 & \teamcolor{OvGUMEMoRIAL} OvGUMEMoRIAL & 43.15 $\pm$ 13.88 & 0.85 $\pm$ 0.16 & 79.10 $\pm$ 29.51 & 5x$10^3$ $\pm$ 5x$10^4$ & 12.07 $\pm$ 23.83 & 5.22 $\pm$ 12.43 & 73.00 $\pm$ 21.83 & 74.09 $\pm$ 52.44 & 22.16 $\pm$ 26.82 \\
 & \teamcolor{IITKGP-KLIV} IITKGP-KLIV \vspace{0.5mm}& 35.33 $\pm$ 17.79 & 0.63 $\pm$ 0.36 & 50.14 $\pm$ 46.58 & 13.51 $\pm$ 20.33 & 15.17 $\pm$ 27.32 & 16.69 $\pm$ 19.87 & 40.46 $\pm$ 38.26 & 130.3 $\pm$ 67.59 & 8.39 $\pm$ 22.29 \\
\midrule 
\multirow{7}{*}{\rotatebox[origin=c]{90}{\parbox{0.09\textwidth }{\centering \textbf{Task 5}}}} & \teamcolor{nnU-Net} nnU-Net & \textbf{72.44 $\pm$ 5.05} & 0.95 $\pm$ 0.02 & \textbf{94.6 $\pm$ 1.59} & 5.07 $\pm$ 2.57 & \textbf{37.17 $\pm$ 20.83} & 1.05 $\pm$ 0.55 & \textbf{92.98 $\pm$ 3.69} & 14.87 $\pm$ 5.88 & \textbf{75.52 $\pm$ 8.83} \\
 & \teamcolor{PKDIA} PKDIA & 66.46 $\pm$ 5.81 & 0.93 $\pm$ 0.02 & 92.97 $\pm$ 1.78 & 6.91 $\pm$ 3.27 & 28.65 $\pm$ 18.05 & 1.43 $\pm$ 0.59 & 90.44 $\pm$ 3.96 & 20.1 $\pm$ 5.90 & 66.71 $\pm$ 9.38 \\
 & \teamcolor{mountain} Mountain & 60.2 $\pm$ 8.69 & 0.90 $\pm$ 0.03 & 85.81 $\pm$ 10.18 & 8.04 $\pm$ 3.97 & 21.53 $\pm$ 15.50 & 2.27 $\pm$ 0.92 & 84.85 $\pm$ 6.11 & 25.57 $\pm$ 8.42 & 58.66 $\pm$ 10.81 \\
 & \teamcolor{ISDUE} ISDUE & 56.25 $\pm$ 19.63 & 0.83 $\pm$ 0.23 & 79.52 $\pm$ 28.07 & 18.33 $\pm$ 27.58 & 12.51 $\pm$ 15.14 & 5.82 $\pm$ 11.72 & 77.88 $\pm$ 26.93 & 32.88 $\pm$ 33.38 & 57.05 $\pm$ 21.46 \\
 & \teamcolor{METU_MMLAB} METU\_MMLAB & 56.01 $\pm$ 6.79 & 0.89 $\pm$ 0.03 & 80.22 $\pm$ 12.37 & 12.44 $\pm$ 4.99 & 15.63 $\pm$ 13.93 & 3.21 $\pm$ 1.39 & 79.19 $\pm$ 8.01 & 32.70 $\pm$ 9.65 & 49.29 $\pm$ 12.69 \\
 & \teamcolor{OvGUMEMoRIAL} OvGUMEMoRIAL & 44.34 $\pm$ 14.92 & 0.79 $\pm$ 0.15 & 64.37 $\pm$ 32.19 & 76.64 $\pm$ 122.44 & 9.45 $\pm$ 11.98 & 4.56 $\pm$ 3.15 & 71.11 $\pm$ 18.22 & 42.93 $\pm$ 17.86 & 39.48 $\pm$ 16.67 \\
 & \teamcolor{IITKGP-KLIV} IITKGP-KLIV & 25.63 $\pm$ 5.64 & 0.56 $\pm$ 0.06 & 41.91 $\pm$ 11.16 & 13.38 $\pm$ 11.2 & 11.74 $\pm$ 11.08 & 18.7 $\pm$ 6.11 & 35.92 $\pm$ 8.71 & 114.51 $\pm$ 45.63 & 11.65 $\pm$ 13.00\\
\bottomrule
\end{tabular}%
}
\raggedright\footnotesize{* Corrected submission of PKDIA right after the on-site session (i.e. During the challenge, they have submitted the same results, but in reversed orientation. Therefore, the winner of Task 2 at conference session is the MedianCHAOS6).}
\end{table*}

The training dataset was published approximately three months before the on-site session. The test dataset was given 24 hours before the challenge session. The submissions were evaluated during the conference, and the winners were announced. After the on-site session, training and test datasets were published on the zenodo.org website \citep{Kavur2019} and the online submission system was activated on the challenge website.

To compare the automatic DL methods with semi-automatic ones, interactive methods including both traditional iterative models and more recent techniques \lk{were} employed from our previous work \citep{kavur2019b}. In this respect, we report the results and discuss the accuracy and repeatability of emerging automatic DL algorithms with those of well-established interactive methods, which are applied by a team of imaging scientists and radiologists through two dedicated viewers: Slicer \citep{Kikinis2014} and exploreDICOM \citep{Fischer2010}.

There exist two separate leaderboards at the challenge website, one for the conference session\footnote{https://chaos.grand-challenge.org/Results\_CHAOS/} and another for post-conference online submissions\footnote{https://chaos.grand-challenge.org/evaluation/results/}. Detailed metric values and converted scores are presented in Tab.\ref{tab:scores}. Box plots of all results for each task are presented separately in Fig.\ref{fig:task_mean}. Also, scores on each testing case are shown in Fig.\ref{fig:task_all} for all tasks. As expected, the tasks that received the highest number of submissions and scores were the ones focusing on the segmentation of a single organ from a single modality. Thus, the vast majority of the submissions were for liver segmentation from CT images (Task 2), followed by liver segmentation from MR images (Task 3). Accordingly, in the following subsections, the results are presented in the order of performance/participation in Tab.\ref{tab:submissions} (i.e. from the task having the \lk{most} submissions to the one having the \lk{fewest}).  In this way,  the segmentation from cross- and multi-modality/organ concepts (Tasks 1 and 4) are discussed in light of the performances \lk{of} more conventional approaches (Tasks 2, 3, and 5). 

\subsection{Remarks about Multiple Submissions}

\begin{figure*}[!ht]
    \centering
        \includegraphics[width=1.0\textwidth]{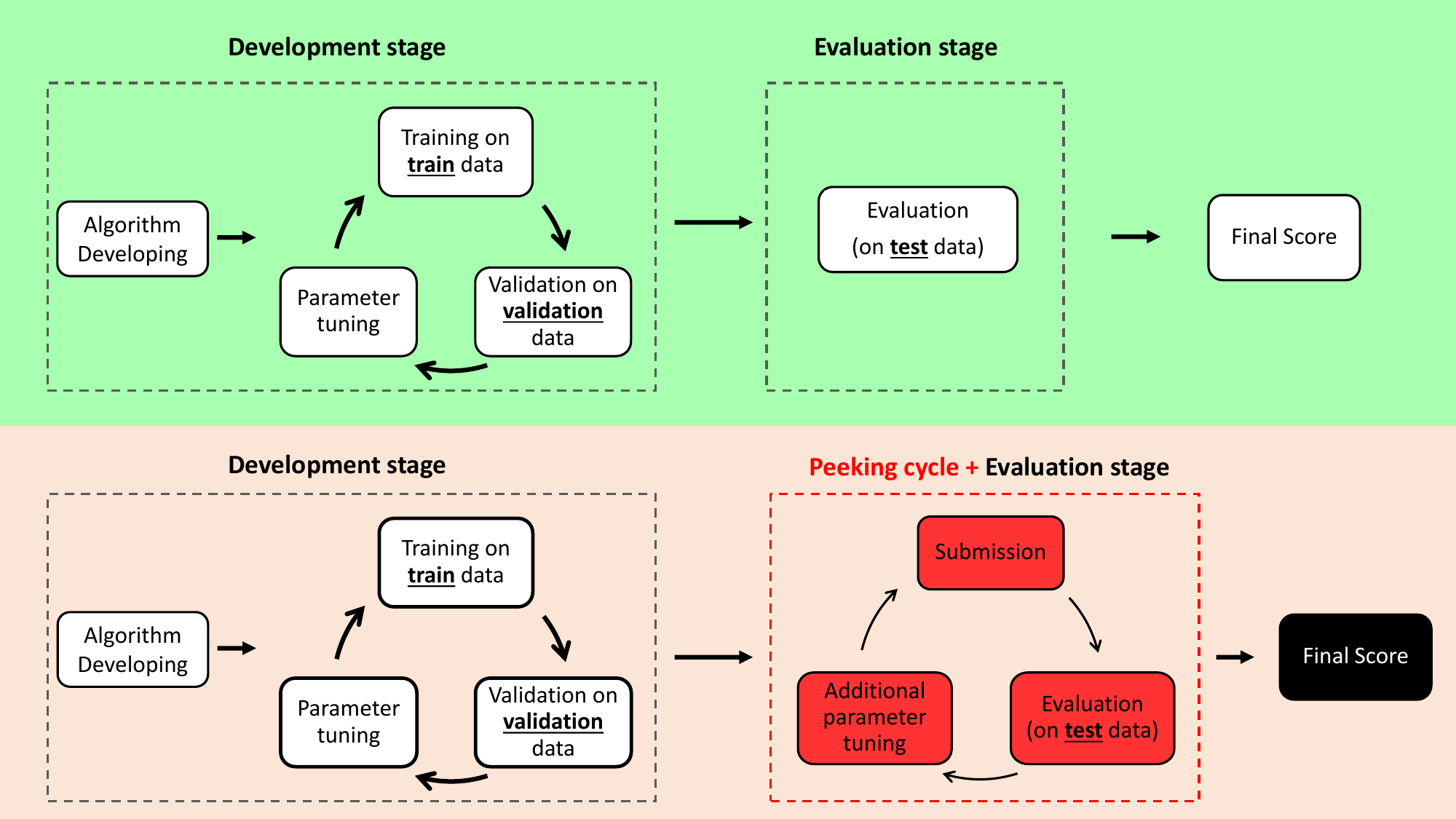}
        \caption{Illustration of a fair study (green area) and peeking attempts (red area)}
        \label{fig:peeking}
\end{figure*}

Currently, more than 1500 participants are registered to the CHAOS challenge through the “chaos.grand-challenge.org” website. There is no direct correlation between “the number of participants” and “the number of submitted results” (i.e. $550$), because there are passive participants, who never submitted a result, \lk{as well as} very active ones, who \lk{made} multiple submissions. The organizers put no restrictions for registration as long as the candidates agree to the terms of use (i.e. research purpose only etc.). This first step of joining is intentionally left unrestricted to encourage \lk{wide participation}. Moreover, there is also no restriction \lk{requiring the participant} to submit a result. The registration and evaluation system is completely automated (thanks to the grand-challenge.org website design) and all evaluated results are immediately published on the leaderboard regardless of their score. There is no disqualification or exclusion based on the number of submissions. However, the organizers \lk{check the submitted results} quantitatively and qualitatively to prevent \textit{peeking} as described in the following paragraphs.

\emre{Although test datasets should only be considered as the unseen (new) data provided to the algorithms to evaluate their performance, there is a way to use them at the algorithm development stage. This kind of use is called ``peeking'', which is done through reporting too many performance results by iterative submissions \citep{Kuncheva2014}. We claim that peeking can be considered as one of the shortcomings of the image segmentation grand-challenges. Since access to the ground truth is not required, peeking makes it possible to use test data to tune parameters, although the parameter tuning needs to be done during the validation phase.
Eventually, the peeking phenomenon is an important issue and a \lk{still-existing} problem, particularly for evaluating the real-life performance of machine learning-based models. Comparison of a fair development and peeking attempts is shown in Fig.\ref{fig:peeking}. After rigorous analysis of literature and the outcomes of the previous challenges, the CHAOS organizers determine potential source of peeking as follows:}

\emre{Multiple submissions are normally allowed, and the results are disclosed to the participant. This allows the participants to tune their model on the test data, which is a form of ‘peeking’ \citep{Smialowski10,Reunanen03, Diciotti13}. Therefore, using the outcomes (performance metrics vs resulting images) of successive submissions to fine-tune a model may be over-tuned on the test data. Besides, at each iteration, the model likely to become more sophisticated and not always reproducible.}


\begin{table}[!b]
\centering
\caption{Results of selected peeking attempts that have been obtained from online results of CHAOS challenge. The impact of peeking can be observed from score changes. (Team names were anonymized.)}
\label{tab:peeking_efffects}
\def\arraystretch{1.3}
\begin{tabular}{l C{3cm}C{2cm}} \toprule
Participant & Number of iterative submissions & Score change \\ \midrule
Team A & 21 & +29.09\% \\
Team B & 19 & +15.71\% \\
Team C & 16 & +12.47\% \\
Team D & 15 & +30.02\% \\
Team E & 15 & +26.06\% \\
Team F & 13 & +10.10\% \\
\bottomrule
\end{tabular}
\end{table}

\emre{This issue requires further and in-depth analysis through extensive experimentation. During the evaluation of the results submitted to CHAOS (besides allowing multiple submissions without any restrictions) the results are also analyzed for possible peeking attempts: Related observations are discussed below. Since there is still no statistical tool that can mathematically prove peeking through multiple submissions, the corresponding analyses are carried out interactively by communicating with the participants.}


\emre{To differentiate the reason behind a score increase (i.e. due to the advancement of the technique/model or due to peeking), we consider the indication of peeking to two cases: 1) Submitting consecutive results in short time intervals, and 2) Having only data specific changes at succeeding submission (i.e. changes only at some particular cases). Although this is clearly a limited subset of all cases, the available tools only allow the analysis of these conditions. If the above-mentioned suspicions are supported by evaluation metrics, then the participants are asked to justify the improvements leading to their better performance. An example of a suspicious condition can be described as follows:  }
  
\emre{Assume that a participating Team X submitted their results and received Score $Y$. Just, a couple of hours later, they have submitted another result and received score $Y+\alpha$. Of course, this can be attributed to model improvement or parameter adjustment, but when their results are investigated case by case, it was observed that only the performance of some particular cases was changed. If the team provides no reasonable explanation for this improvement, then corresponding participants were assumed to benefit from peeking. The number of submissions and the percentage of score increase of the detected teams is given in Tab.\ref{tab:peeking_efffects}. These results show that the impact of the peeking can be noteworthy in some cases.}

\subsection{CT Liver Segmentation (Task 2)}

\begin{figure*}[!t]
    \centering
        \includegraphics[width=1.0\textwidth]{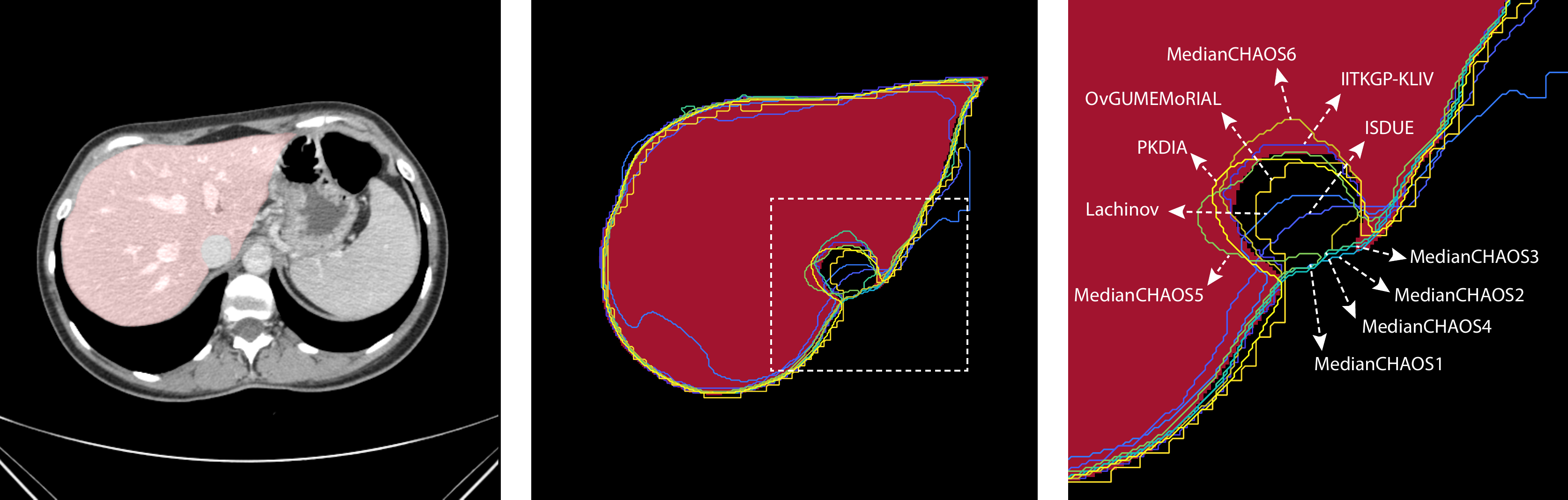}
        \caption{Example image from the CHAOS CT dataset, (case 35, slice 95), borders of segmentation results on ground truth mask and zoomed onto inferior vena cava (IVC) region (marked with dashed lines on the middle image). In this example, the contrast between liver tissue and IVC is relatively lower due to sub-optimal timing during the CT scan. Accordingly, it creates a challenging case for the the participating algorithms. (Scores of this slice are; PKDIA:91.13, MedianCHAOS6:91.84, MedianCHAOS3:85.42, MedianCHAOS1:82.55, MedianCHAOS2:83.46, MedianCHAOS5:88.38, OvGUMEMoRIAL:85.61, MedianCHAOS4:81.74, ISDUE:65.49, IITKGP-KLIV:66.27, Lachinov:64.18)}
\label{fig:task2_liver} 
\end{figure*}

This task includes one of the most \emre{frequently} studied cases and a very mature field of abdominal segmentation. Therefore, it provides a good opportunity to test the effectiveness of the participating models compared to the existing approaches. Although the provided datasets only include healthy organs, the injection of contrast media creates several additional challenges, as described in Section III.B.  Nevertheless, the highest scores of the challenge were obtained in this task (Fig.\ref{fig:task_mean}b). 

The on-site winner was MedianCHAOS with a score of 80.45$\pm$8.61 and the online winner is PKDIA with 82.46$\pm$8.47. Being an ensemble strategy, the performance of the sub-networks of MedianCHAOS is illustrated in Fig. \ref{fig:task_all}.c. When individual metrics are analyzed, DICE performance seem to be outstanding (i.e. 0.98$\pm$0.00) for both winners (i.e. scores 97.79$\pm$0.43 for PKDIA and 97.55$\pm$0.42 for MedianCHAOS). Similarly, ASSD performances have very high mean and small variance (i.e. 0.89$\pm$0.36 [score: 94.06$\pm$2.37] for PKDIA and  0.90$\pm$0.24 [94.02$\pm$1.6] for MedianCHAOS). On the other hand, RAVD and MSSD scores are \emre{dramatically} low resulting in reduced overall performance. Actually, this outcome is valid for all tasks and participating methods. 

Regarding semi-automatic approaches in \citep{kavur2019b}, the best three \lk{entries} received scores \emre{of} 72.8 (active contours with a mean interaction time (MIT) of 25 minutes ), 68.1 (robust static segmenter having an MIT of 17 minutes), and 62.3 (i.e. watershed with MIT of 8 minutes). Thus, the successful designs among participants in deep learning-based automatic segmentation algorithms have outperformed  the interactive approaches by a large margin. \lk{The quality of the segmentation} reaches almost the inter-expert level for volumetric analysis and average surface differences. However, there is still a need for improvement considering the metrics related to maximum error margins (i.e. RAVD and MSSD). An important drawback of the deep approaches is \lk{that} they might completely fail and generate unreasonably low scores for particular cases, such as the inferior vena cava region shown in Fig.\ref{fig:task2_liver}.

Regarding the effect of architectural design on performance, comparative analyses have been performed through some well-established deep frameworks (i.e. DeepMedic \citep{Kamnitsas_2017} and NiftyNet \citep{GIBSON2018113}). These models have been applied with their default parameters and they have both achieved scores of around 70. Thus, considering the participating models that have received \lk{scores} below 70, it is safe to conclude that, 
\lk{ crafting} the new deep architectural designs \lk{or extensive} parameter \lk{tuning do} not necessarily translate into more successful systems.  

\subsection{MR Liver Segmentation (Task 3)}
Segmentation from MR can be considered a more difficult operation compared to segmentation from CT because CT images have a typical histogram and dynamic range defined by Hounsfield Units~(HU), whereas MRI does not have such a standardization. Moreover, artifacts and other factors in clinical routine cause critical degradation of the MR image quality. The on-site winner of this task is PKDIA with a score of 70.71$\pm$6.40. PKDIA had the most successful results not only for the mean score but also for the distribution of the results (shown in Fig.\ref{fig:task_mean}c and \ref{fig:task_all}d). Robustness to the deviations in MR data quality is an important factor that affects performance. For instance, CIR\_MPerkonigg, which has the most successful scores for some cases, could not show a high overall score.

The online winner is nnU-Net with 75.10$\pm$7.61. When the scores of individual metrics are analyzed for PKDIA and nnU-Net, DICE (i.e. 0.94$\pm$0.01 [score: 94.47$\pm$1.38] for PKDIA and 0.95$\pm$0.01 [score: 95.42$\pm$1.32] for nnU-Net) and ASSD  (i.e. 1.32$\pm$0.83 [score: 91.19$\pm$5.55] for nnU-Net and  1.56$\pm$0.68 [score: 89.58$\pm$4.54] for PKDIA) performance \emre{is} again extremely good, while RAVD and MSSD scores are critically lower than the CT results. The reason behind this can also be attributed to the lower resolution and higher spacing of the MR data, which cause a higher spatial error for each misclassified pixel/voxel (see Tab.\ref{tab:data_stats}). Comparisons with the interactive methods show that they tend to make regional mistakes due to the spatial enlargement strategies. The main challenge for them is to differentiate the outline when the liver is adjacent to isodense structures. On the other hand, automatic methods show much more distributed mistakes all over the liver. Further analysis also revealed that interactive segmentation methods tend to make fewer over-segmentations. This is partly related to iterative parameter adjustment \lk{by} the operator which prevents unexpected results. Overall, the participating methods performed equally well with  interactive methods when only volumetric metrics are considered. However, the interaction seems to outperform deep models for other metrics. 

\subsection{CT-MR Liver (Cross-Modality) Segmentation (Task 1)}
This task targets cross-modality learning and it involves the usage of CT and MR information together during training. A model that can effectively accomplish cross-modality learning would:  1) help to satisfy large amounts of training data by providing more images and 2) reveal common features of incorporated modalities for an organ. To compare cross-modality learning with individual ones,  Fig.\ref{fig:task_all}a should be compared to  Fig.\ref{fig:task_all}c for CT. Such a comparison clearly reveals that participating models trained only on CT data show obviously better performance than models trained on both modalities. A similar observation can also be made for MR results by observing Fig.\ref{fig:task_all}b and  Fig.\ref{fig:task_all}d. \lk{This shows that there are still improvements necessary for a single solution working on images of multiple modalities. However, remarkable developments in the machine learning field may overcome these problems.}

\begin{figure*}[!b]
    \centering
        \includegraphics[width=0.99\textwidth]{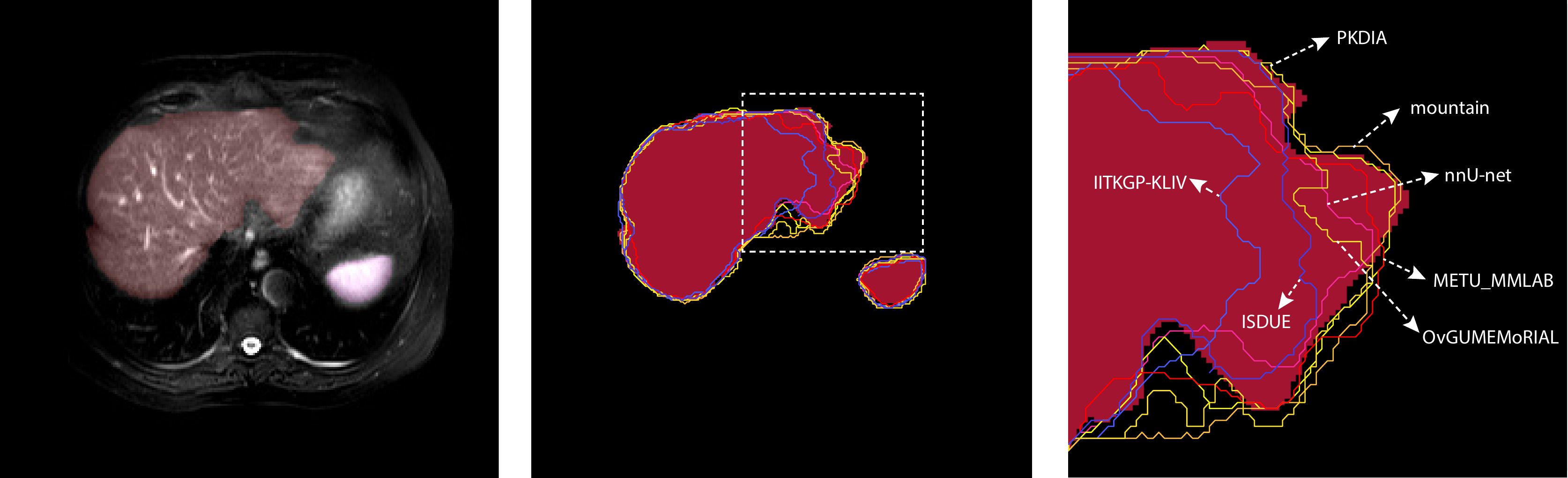}
         \caption{Example image from the CHAOS MRI T2SPIR dataset, (case 24, slice 23), borders of segmentation results on ground truth mask for liver and spleen, and zoomed onto the marked region. The inhomogeneous intensity the distribution of the liver cause errors (both over- and under-segmentation in specific regions) on the segmentation results. In general, such a problem is not observed for the segmentation of kidneys and spleen. (Scores of this slice are; nnU-net:62.07, PKDIA:65.15, mountain:59.08, METU\_MMLAB:57.98, ISDUE:53.76, OvGUMEMoRIAL:49.65, IITKGP-KLIV:23.19) }
\label{fig:task5_liver} 
\end{figure*}

The on-site winner of this task was OvGUMEMoRIAL with a score of 55.78$\pm$19.20. Although its DICE performance is quite satisfactory (i.e. 0.88$\pm$0.15, corresponding to a score of 83.14$\pm$0.43), the other measures cause the low grade. Here, a very interesting observation is that the score of OvGUMEMoRIAL is lower than its score on CT (61.13$\pm$19.72) but higher than  MR (41.15$\pm$21.61). Another interesting observation of the highest-scoring non-ensemble model, PKDIA, both for Task 2 (CT) and Task 1 (MR), had a dramatic performance drop in this task. 

\begin{figure*}[!t]
    \centering
        \includegraphics[width=0.90\textwidth]{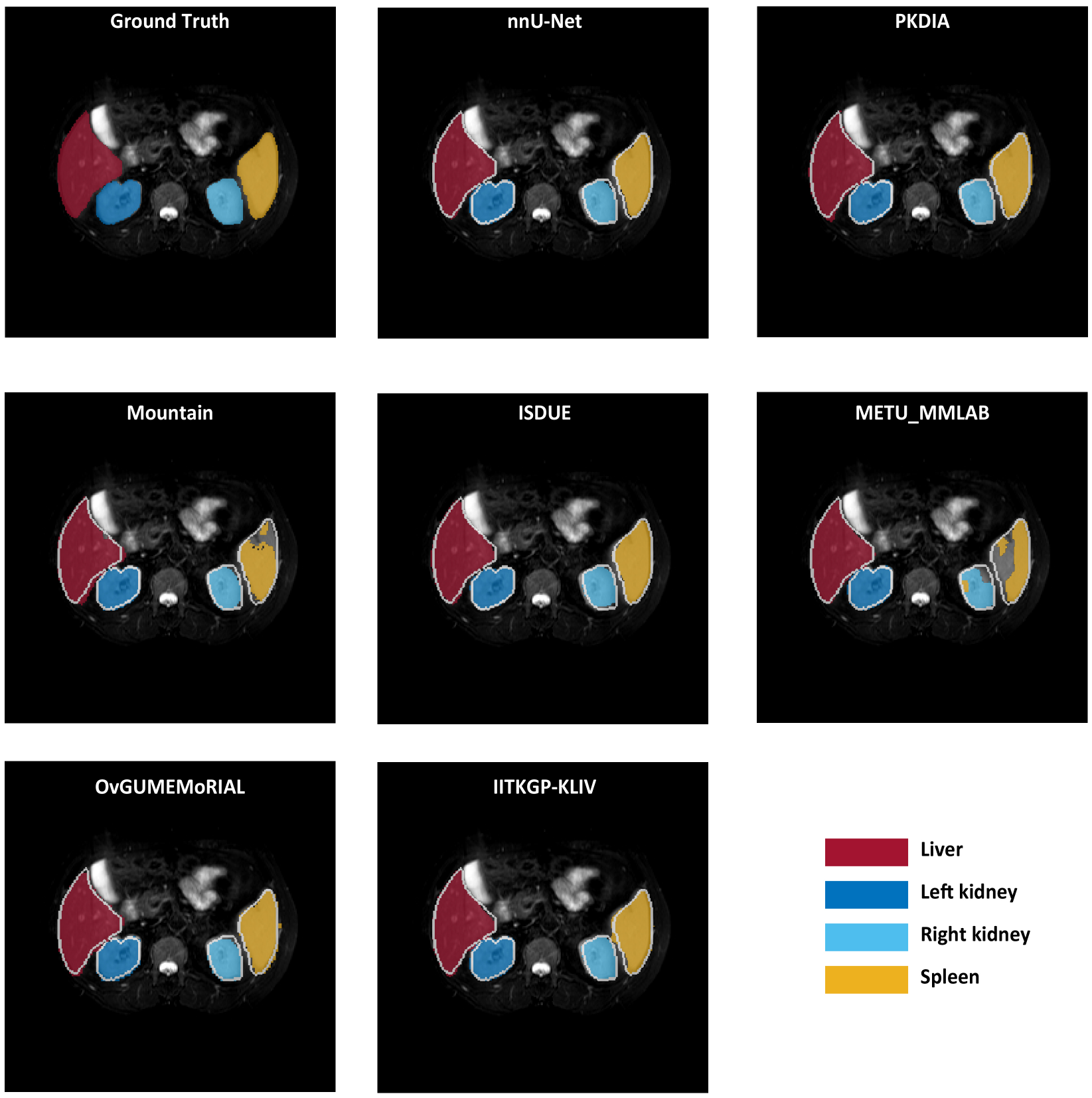}
        \caption{Illustration of ground truth and all results for Task 5. The image was taken from the CHAOS MR dataset (case 40, slice 15). White lines on the results represent borders of ground truth. (Scores of this slice are; nnU-net:74.52, PKDIA:74.37, mountain:44.09, METU\_MMLAB:42.32, ISDUE:60.90, OvGUMEMoRIAL:55.21, IITKGP-KLIV:55.46) }
\label{fig:task5} 
\end{figure*}

It is important to examine the scores of cases with their distribution across all data. This can help to analyze the generalization capabilities and real-life use of these systems. For example, Fig.\ref{fig:task_mean}.a shows a noteworthy situation. The winner of Task 1, OvGUMEMoRIAL, shows lower performance than the second method (ISDUE) in terms of standard deviation. Fig.\ref{fig:task_all}a and \ref{fig:task_all}b show that the competing algorithms have slightly higher scores on the CT data than on the MR data. However, if we consider the scattering of the individual scores along with the data, CT scores have higher variability. This shows that reaching equal generalization for multiple modalities is a challenging task for Convolutional Neural Networks~(CNNs).

\subsection{Multi-Modal MR Abdominal Organ Segmentation (Task 5)}

Task 5 investigates how DL models contribute to the development of more comprehensive computational anatomical models leading to multi-organ related tasks. Deep models have the potential to provide a complete representation of the complex and flexible abdominal anatomy by incorporating inter-organ relations through their internal hierarchical feature extraction processes. In order to qualitatively analyze their performance, an illustration of ground truth and results of all teams on a sample image was presented in Fig.\ref{fig:task5_liver} and \ref{fig:task5}.

The on-site winner was PKDIA with a score of 66.46$\pm$0.81 and the online winner is nnU-Net with 72.44$\pm$5.05. When the scores of individual metrics are analyzed in comparison to Task 3, the DICE performance seems to remain almost the same for nnU-Net and PKDIA. This is an \emre{important} outcome as all four organs are segmented instead of a single one. It is also worth to point \emre{out} that the model of the third-place (i.e. Mountain) has almost exactly the same overall score for Task 3 and 5. The same observation is also valid for the standard deviation of these models. Considering RAVD, the performance decrease seems to be higher compared to DICE. These reduced DICE and RAVD performance \emre{is} partially compensated by better MSSD and ASSD performances. 

Follow-up studies that use the CHAOS dataset, but their outputs were not submitted in the challenge have also reported results \citep{Sinha2020} for this task. Considering an attention-based model \citep{wang2018} as the baseline (DICE: 0.83 $\pm$ 0.06), an ablation study is carried out that reported an increased segmentation performance \citep{Sinha2020}. The limitations \emre{are reduced} by capturing richer contextual dependencies through guided self-attention mechanisms. Architectural modifications for integrating local features with their global dependencies and adaptive highlighting of interdependent channel maps reveal better performance (DICE: 0.87 $\pm$ 0.05) compared to some other models such as U-net \citep{ronneberger2015u} (DICE: 0.81 $\pm$ 0.08), DANet \citep{Fu2019} (DICE: 0.83 $\pm$ 0.10), PAN(ResNet) \citep{Li2018} (DICE: 0.84 $\pm$ 0.06), and UNet Attention \citep{Schlemper_2019} (DICE: 0.85 $\pm$ 0.05).

Despite these slight improvements \lk{achieved} by novel architectural designs, the performance of the proposed models still seems to below the best three contestants (i.e. nnUnet-0.95 $\pm$ 0.02, PKDIA-0.93 $\pm$ 0.02 and Mountain- 0.90 $\pm$ 0.03) of the CHAOS challenge. This is also observed for other metrics such as ASSD (OvGUMEMoRIAL: 1.05 $\pm$ 0.55). Nevertheless, the modifications \lk{by} \citep{Sinha2020} reduced the standard deviation of the other metrics rather than their mean values. The qualitative analysis performed to visualize the effect of the proposed modifications illustrate that some models (such as UNet) typically under-segments certain organs, produce smoother segmentations causing loss of fine-grained details. The architectural modifications are especially helpful to compensate for such drawbacks by focusing the attention of the model on anatomically more relevant areas.

\subsection{CT-MR Abdominal Organ Segmentation (Cross-Modality Multi Modal) (Task 4)}
This task covers the segmentation of both the  liver in CT and four abdominal organs in MRI data. Hence, it can be considered as the most difficult task since it contains both cross-modality learning and multiple organ segmentation. Therefore, it is not surprising that it has the lowest \emre{participation} and scores. 

The on-site winner was ISDUE with a score of 58.69$\pm$18.65. Fig. \ref{fig:task_all}.e-f shows that their solution had consistent and high-performance distribution in both CT and MR data. It can be thought that two convolutional encoders in their system boost performance on cross-modality data. These encoders are able to compress information about anatomy. On the other hand, PKDIA also shows promising performance with a score of 49.63$\pm$23.25. Despite their success on MRI sets, the CT performance can be considered unsatisfactory, similar to their situation at Task 1. This reveals that the CNN-based encoder may not be trained \emre{effectively}. As the encoder part of their solution relies on transfer learning, fine-tuning the pre-trained weights was not successful in multiple modalities. The OvGUMEMoRIAL team achieved the third position with an average score of 43.15 and they have a balanced performance on both modalities. Their method can be considered successful in terms of generalization, \lk{compared to the other} participating teams.

Together with the outcomes of Task 1 and 5, it is shown that in \lk{the} current strategies and architectures, CNNs have better segmentation performance on single modality tasks. This might be considered as an expected outcome because the success of CNNs is \lk{highly} dependent on the consistency and homogeneity of the data. Using multiple modalities creates a high variance in the data even though all data were normalized. On the other hand, the results also revealed that CNNs have good potential for cross-modality tasks if appropriately extended models are constructed. This potential was not that clear before the development of deep learning strategies for segmentation.

\emre{\subsection{Remarks on Ranking Stability and Robustness}}

\begin{table*}[!t]
\centering
\caption{Mean scores of the submissions against different thresholds. The left column represents the scores with selected thresholds in the challenge. The middle column shows the scores obtained with $5\%$ higher thresholds favoring more precise segmentation results (higher for DICE and lower for other metrics). The right column shows the scores obtained with thresholds reduced by $5\%$ (lower for DICE and higher for other metrics). The best results are given in bold.}
\label{tab:thresholds}
\def\arraystretch{1.3}
\begin{tabular}{clC{4.2cm}C{4.2cm}C{4.2cm}}
\toprule 
 & \hspace{5mm} \textbf{Team Name} & \textbf{Mean score with \newline default thresholds} & \textbf{Mean score with \newline 5\% more precise thresholds} & \textbf{Mean score with \newline 5\% less precise thresholds} \\ \midrule
\multirow{6}{*}{\rotatebox[origin=c]{90}{\parbox{0.1\textwidth }{\centering \textbf{Task 1}}}}  & \teamcolor{OvGUMEMoRIAL} OvGUMEMoRIAL & \textbf{55.78 $\pm$ 19.20}& \textbf{54.36 $\pm$ 19.98} & \textbf{57.04 $\pm$ 18.76} \\
 & \teamcolor{ISDUE} ISDUE & 55.48 $\pm$ 16.59 & 51.76 $\pm$ 17.02 & 56.64 $\pm$ 16.24 \\
 & \teamcolor{PKDIA} PKDIA & 50.66 $\pm$ 23.95 & 46.61 $\pm$ 24.12 & 52.14 $\pm$ 23.02  \\
 & \teamcolor{lachinov} Lachinov & 45.10 $\pm$ 21.91 & 44.85 $\pm$ 22.14 & 48.09 $\pm$ 20.85  \\
 & \teamcolor{METU_MMLAB} METU\_MMLAB & 42.54 $\pm$ 18.79 & 38.54 $\pm$ 17.98 & 42.25 $\pm$ 18.22  \\
 & \teamcolor{IITKGP-KLIV} IITKGP-KLIV \vspace{0.5mm} & 40.34 $\pm$ 20.25 & 34.65 $\pm$ 20.95 & 43.21 $\pm$ 19.87  \\
\midrule 
\multirow{11}{*}{\rotatebox[origin=c]{90}{\parbox{0.1\textwidth }{\centering \textbf{Task 2}}}} & \teamcolor{PKDIA} PKDIA & \textbf{82.46 $\pm$ 8.47} & \textbf{81.83 $\pm$ 8.11}  & \textbf{83.51 $\pm$ 8.58} \\
 & \teamcolor{MedianCHAOS6} MedianCHAOS6 & 80.45 $\pm$ 8.61 & 79.57 $\pm$ 8.17 & 81.28 $\pm$ 8.98 \\
 & \teamcolor{MedianCHAOS3} MedianCHAOS3 & 80.43 $\pm$ 9.23 & 79.55 $\pm$ 9.88 & 81.27 $\pm$ 8.99 \\
 & \teamcolor{MedianCHAOS1} MedianCHAOS1 & 79.91 $\pm$ 9.76 & 78.95 $\pm$ 10.01 & 80.77  $\pm$ 9.25 \\
 & \teamcolor{MedianCHAOS2} MedianCHAOS2 & 79.78 $\pm$ 9.68 & 78.86 $\pm$ 10.09 & 80.65 $\pm$ 9.23 \\
 & \teamcolor{MedianCHAOS5} MedianCHAOS5 & 73.39 $\pm$ 6.96 & 72.35 $\pm$ 7.24 & 74.32 $\pm$ 6.25 \\
 & \teamcolor{OvGUMEMoRIAL} OvGUMEMoRIAL & 61.13 $\pm$ 19.72 & 60.00 $\pm$ 20.14 & 62.16 $\pm$ 19.11 \\
 & \teamcolor{MedianCHAOS4} MedianCHAOS4 & 59.05 $\pm$ 16.00 & 58.14 $\pm$ 16.88 & 59.88 $\pm$ 15.24 \\
 & \teamcolor{ISDUE} ISDUE & 55.79 $\pm$ 11.91 & 54.87 $\pm$ 12.25 & 57.61 $\pm$ 11.43 \\
 & \teamcolor{IITKGP-KLIV} IITKGP-KLIV & 55.35 $\pm$ 17.58 & 54.15 $\pm$ 18.27 & 56.48 $\pm$ 17.26 \\
 & \teamcolor{lachinov} Lachinov & 39.86 $\pm$ 27.90 & 36.75 $\pm$ 28.14 & 41.94 $\pm$ 27.52 \\
\midrule 
\multirow{9}{*}{\rotatebox[origin=c]{90}{\parbox{0.1\textwidth }{\centering \textbf{Task 3}}}} & \teamcolor{nnU-Net} nnU-Net & \textbf{75.10 $\pm$ 7.61} & \textbf{74.65 $\pm$ 7.89} & \textbf{76.85 $\pm$ 7.11} \\
 & \teamcolor{PKDIA} PKDIA & 70.71 $\pm$ 6.40 & 69.59 $\pm$ 6.95 & 71.76 $\pm$ 6.02 \\
 & \teamcolor{mountain} Mountain & 60.82 $\pm$ 10.94 & 59.18 $\pm$ 11.24 & 61.95 $\pm$ 10.56 \\
 & \teamcolor{ISDUE} ISDUE & 55.17 $\pm$ 20.57 & 54.11 $\pm$ 21.01 & 56.15 $\pm$ 20.06 \\
 & \teamcolor{CIR-MPerkonigg} CIR\_MPerkonigg & 53.60 $\pm$ 17.92 & 52.71 $\pm$ 18.31 & 55.45 $\pm$ 17.16 \\
 & \teamcolor{METU_MMLAB} METU\_MMLAB & 53.15 $\pm$ 10.92 & 52.05 $\pm$ 11.17 & 54.32 $\pm$ 10.33 \\
 & \teamcolor{lachinov} Lachinov & 50.34 $\pm$ 12.22 & 48.91 $\pm$ 12.89 & 51.17 $\pm$ 12.03 \\
 & \teamcolor{OvGUMEMoRIAL} OvGUMEMoRIAL & 41.15 $\pm$ 21.61 & 39.07 $\pm$ 22.10 & 42.60 $\pm$ 21.35 \\
 & \teamcolor{IITKGP-KLIV} IITKGP-KLIV \vspace{0.5mm}& 34.69 $\pm$ 8.49 & 33.96 $\pm$ 8.98 & 35.40 $\pm$ 8.12 \\
\midrule 
\multirow{4}{*}{\rotatebox[origin=c]{90}{\parbox{0.09\textwidth }{\centering \textbf{Task 4}}}} & \teamcolor{ISDUE} ISDUE & \textbf{58.69 $\pm$ 18.65} & \textbf{56.27 $\pm$ 18.97} & \textbf{58.79 $\pm$ 18.12} \\
 & \teamcolor{PKDIA} PKDIA & 49.63 $\pm$ 23.25 & 48.85 $\pm$ 23.96 & 50.65 $\pm$ 22.99 \\
 & \teamcolor{OvGUMEMoRIAL} OvGUMEMoRIAL & 43.15 $\pm$ 13.88 & 43.99 $\pm$ 14.02 & 48.18 $\pm$ 13.56 \\
 & \teamcolor{IITKGP-KLIV} IITKGP-KLIV \vspace{0.5mm}& 35.33 $\pm$ 17.79 & 24.61 $\pm$ 18.02 & 37.36 $\pm$ 17.32 \\
\midrule 
\multirow{7}{*}{\rotatebox[origin=c]{90}{\parbox{0.09\textwidth }{\centering \textbf{Task 5}}}} & \teamcolor{nnU-Net} nnU-Net & \textbf{72.44 $\pm$ 5.05} & \textbf{71.89 $\pm$ 5.56} & \textbf{73.57 $\pm$ 4.98} \\
 & \teamcolor{PKDIA} PKDIA & 66.46 $\pm$ 5.81 & 64.77 $\pm$ 6.01 & 67.32 $\pm$ 5.45 \\
 & \teamcolor{mountain} Mountain & 60.20 $\pm$ 8.69 & 57.72 $\pm$ 8.98 & 61.54 $\pm$ 8.25 \\
 & \teamcolor{ISDUE} ISDUE & 56.25 $\pm$ 19.63 & 54.78 $\pm$ 20.21 & 57.51 $\pm$ 19.56 \\
 & \teamcolor{METU_MMLAB} METU\_MMLAB & 56.01 $\pm$ 6.79 & 52.73 $\pm$ 7.12 & 57.95 $\pm$ 6.23 \\
 & \teamcolor{OvGUMEMoRIAL} OvGUMEMoRIAL & 44.34 $\pm$ 14.92 & 40.97 $\pm$ 15.25 & 46.07 $\pm$ 14.65 \\
 & \teamcolor{IITKGP-KLIV} IITKGP-KLIV & 25.63 $\pm$ 5.64 & 24.41 $\pm$ 5.89 & 26.66 $\pm$ 5.38 \\
\bottomrule
\end{tabular}%
\end{table*}

It is well known and extensively discussed in the medical imaging community that the evaluation strategy plays a key role in the rankings  ~\citep{Maier-Hein2018}. The performance of a model relies on how the metrics are transformed into the scores and the main factor at this transformation is the selection of the thresholds. In CHAOS, the expert physicians (i.e. a team of radiologists and surgeons) determined these thresholds after extensive discussions. Nevertheless, to explore the stability of the rankings through their dependency on threshold, scores are re-calculated using new threshold values. To measure the sensitivity, the threshold \lk{were} changed by 5\% in two ways: 

1. Thresholds were increased by 5\% for accepting/favoring more precise segmentation results (higher for DICE, lower for the other metrics). The DICE threshold is increased to 0.84 while RAVD, ASSD, and MSSD were decreased to 4.75\%, 14.2mm, and 57mm respectively.

2. Thresholds \lk{were} decreased by 5\% for accepting/favoring less precise segmentation results (lower for DICE, higher for the other metrics) These values \lk{were} calculated as 0.76 for DICE, 5.25\% for RAVD, 15.75mm for ASSD, and 63mm for MSSD.  

For both cases, all of the scores and rankings were re-calculated. The results are presented in Table 11, which shows that there are no noteworthy changes in the rankings. The only change is between teams METU\_MMLAB and ISDUE (4th and 5th places on the scoreboard) in Task 5. However, the scores of these teams are very close and the change occurs in the decimals (i.e. 0.24 difference in favor of ISDUE at original thresholds and 0.44 in favor of METU\_MMLAB when thresholds are decreased by 5\%). According to these results, it is possible to state that the rankings in CHAOS are robust and they are not influenced by small threshold changes.

The authors believe that the stability of the rankings is achieved by carefully following the organization suggestions given at ~\citep{Maier-Hein2018}. \lk{As highlighted by \cite{Maier-Hein2018}, when designing the challenge (especially for evaluation stage), requirements of three main points that can strongly affect scores are satisfied in the CHAOS challenge:} 

1) Preventing ranking alterations due to minor changes in metrics: CHAOS rankings are shown to be robust to such changes as \lk{shown in} Tab.\ref{tab:thresholds}.

2) Making the ground truth less dependent on annotator differences: This issue is resolved by using three annotators and their consensus as the ground truth

3) \lk{Handling of missing data} to block rank manipulation: This is handled by giving zero points to non-existing cases. 

The remaining major factor in the evaluation, the aggregation of different metrics is chosen to be averaging in CHAOS (instead of alternatives such as median ~\citep{Maier-Hein2018}), since the physicians find the used metrics equally important from multiple clinical perspectives (such as surgical precision, follow-up analysis, etc.)
\newline

\section{Discussions and Conclusion}
In this paper, we presented the CHAOS \lk{challenge}. We generated an unpaired cross-modality (CT-MR), multi-\lk{modality} (MR T1-DUAL in / oppose, T2-SPIR) public dataset for five tasks and evaluated a \emre{considerable} number of newly proposed, well-established, and state-of-the-art segmentation methods. Five different tasks targeting at single modality (CT or MR), cross-modality (CT and MR), and multi-modal (MR T1 in/oppose and T2 sequences) segmentation were prepared. The evaluation is performed using a scoring system based on four metrics. Our results indicate various important outcomes. 

\subsection{Task-Based Conclusions}
Task-based conclusions can be highlighted as follows :
 
1) Since the start of the competition (11 April 2019), the most popular task, Task 2 (liver segmentation on CT), has received more than 200 submissions \lk{within} eight months. Quantitative analyses on Task 2 show that CNNs for segmentation of the liver from CT have achieved a great success. Deep learning-based automatic methods outperformed interactive semi-automatic strategies for CT liver segmentation. They have reached inter-expert variability for DICE and volumetry, but still need some more improvements for distance-based metrics that are critical for determining surgical error margins. Supporting the quantitative analyses, our qualitative observations \lk{suggest that} the top methods can be used in real-life solutions with little efforts on post-processing.  

2) Considering MR liver segmentation (Task 3), the participating deep models have performed almost equally well as interactive ones for DICE, but lack \lk{in} performance for distance-based measures. Given the outstanding results for this task and the fact that the resulting volumes will be visualized by a radiologist-surgeon team prior to various operations in the context of clinical routine, it can be concluded that minimal user interaction, especially in the post-processing phase, would easily bring the single modality MR results to clinically acceptable levels. Of course, this would require not only having a software implementation of the participating methods, but also their integration to an adequate workstation/DICOM viewer, easily accessible in the daily workflow of \lk{the physician.} 

3) 
Deep models perform better in the segmentation of the four abdominal organs (Task 5) compared to the segmentation of only the liver. However, it is not clear whether this improvement can be attributed to multi-tasking. For instance, due to its relatively bigger size and higher shape variations, the MSSD performance of the models for the liver is worse compared to other organs (i.e. MRI average MSSD scores: Liver 61.01 mm, Right Kidney 44.31 mm, Left kidney 46.57 mm, and Spleen  44.22 mm). Accordingly, when all organs are segmented, average MSSD becomes lower (i.e. better) compared to liver segmentation. Our in-depth analyses show that even for slight performance gains, \lk{the} reviewed methods \lk{will need substantial improvement}, or new approaches \lk{have to} be developed. However, the impact and the \emre{importance} of these slight \lk{gains in segmentation quality} may not justify the effort. 

This conclusion is also validated by independent studies, which used the CHAOS dataset and utilized ablation studies to improve model performance. In \citep{Sinha2020}, a series of experiments are performed to validate the individual contribution of different components to the segmentation performance. Compared to the baseline, integrating spatial or attention modules to the architecture \lk{was} observed to increase the performance between 2-3\% for DICE 12-18\% for ASSD while employing both modules only bring slight improvements for DICE and reduce ASSD. Thus, the channel attention module is chosen at the final design simply by observing the parametric simulation results. Besides, such ablation studies relying on extensive experimentation under different settings might cause data-dependent models with \lk{lower} generalization ability.

4) We observed that performances reported for the remaining tasks using cross-modality, i.e., Task1 (Liver Segmentation on CT + MRI) and Task 4 (Segmentation of abdominal organs on CT + MRI), are \emre{clearly} lower than the aforementioned ones.

This shows that despite the \emre{important} developments by DL models for segmentation, \lk{their} application to the real-world clinical problems \lk{still need major progress. }
Thus, cross-modality (CT-MR) learning still proved to be more challenging than individual training. Last but not least, multi-organ cross-modality segmentation remains the most challenging problem until appropriate ways to take advantage of multi-tasking properties of deep models and bigger data advantage of cross-modal medical data are developed. Such complicated tasks could benefit from spatial priors, global topological, or shape-representations in their loss functions as employed by some of the submitted models.

\subsection{Conclusions about Participating Models}

Except for one, all teams involved in this challenge have used a modification of U-Net as a primary regressor model or as a support system. However, the high variance between reported scores 
shows that the understanding of the model performance still relies on many parameters including architectural design, implementation, parametric modifications, optimizations, and tuning. Although several common algorithmic properties can be derived for high-scoring models, an interpretation and/or explanation of why a particular model performs well or not is far from being trivial as relations among these factors are poorly defined. As discussed in the previous challenges, such an analysis is almost impossible on a heterogeneous set of models developed by different teams and programming environments. Moreover, the selection of evaluation metrics, their transformations to scoring, and calculation of the final scores might have an impact on the reported performances.

 \lk{
 We believe DL research is essential to develop effective solutions for medical image segmentation. However, instead of focusing solely on segmentation accuracy, the following issues should be addressed to apply DL methods to real-world clinical use:
 improving generalization through domain adaptation strategies \citep{Junlin2019, Gholami2019, schoenauersebag2019multidomain}, optimizing neural network architectures to reduce the computational cost \citep{Belagiannis2019,Carreira2018}, attaching importance to repeatability and reproducibility \citep{nikolov2018deep}, and focusing on interpretable and \emre{solutions.}}
\emre{Moreover, combining existing strategies, especially atlas based methods (which are still commonly used for benchmarking \citep{Kim2020}), with deep models would enable incorporating spatial knowledge and might have potential to improve performance of DL techniques \citep{GAO2020101831}. }

\subsection{Conclusions about Multiple Submissions, Peeking and Ensembles}

The organizers hope that the submissions included in this article have fair developing stages without peeking attempts. In general, the peeking problem does not exist with on-site challenges that announce the results in a short time. On the other hand, it is a general problem of many online challenges not only in medical image analysis but also in other fields. In CHAOS, various approaches were tested to prevent peeking. Unfortunately, up to our knowledge, there is no clear and elegant way to handle this problem completely. According to our experience from online submissions, the precautions such as limiting the number of submissions, the obligation for using official university/company mail addresses, and demanding a manuscript that explains the methods would be useful, but not perfectly cover all situations. Another alternative viable solution is accepting Docker containers that have source codes of algorithms instead of their results. However, this may need additional preparation time for both challenge organizers and participants.

The effort of participants to outperform other results may lead to misleading performance. The scores obtained at the end of the on-site challenge session makes peeking almost impossible. However, this is not true for online submissions. For this reason, in this paper, we have put a great effort to include online submissions, which not only shows high performance, but also their results can be verified through one of the following ways:

1) Uploading source code and/or model to an open access repository (such as GitHub), 

2) Submitting a PDF document, which explains the utilized approach, and 

3) Providing references that show the previous uses of the utilized method/model. 
 
Finally, it would also be worthwhile to point out that, several medical segmentation challenges demonstrate the ensemble superiority by combining the top-performing models in the scoreboard \citep{Kamnitsas2017b, Isensee2019}. It is well-known that, in many challenges, the amount of training data is limited due to the high expense of gathering and annotating medical volumetric datasets \citep{heimann2009comparison, bilic2019liver, menze2014multimodal}). Being relatively small for proper training of a deep model, this can lead to overfitting of individual models. However, classifier ensembles are known to achieve better results compared to their base classifiers even when those classifiers are over-trained  \citep{Kuncheva2014, prevedello2019challenges}. Accordingly, when the top methods (usually Deep Models) are combined through some rule (such as majority voting), the result usually become better than the best individual result \citep{bilic2019liver, menze2014multimodal,jimenez2016cloud, kavur2019b}. On the other hand, such results should be analyzed carefully due to dependency between the train and test data during the construction of ensembles. 

\section*{Acknowledgments}
The organizers would like to thank Ivana Isgum and Tom Vercauteren in the challenge committee of ISBI 2019 for their guidance and support. We express our gratitude to supporting organizations of the grand-challenge.org platform. We thank Umut Baran Ekinci, Ece Köse, Fabian Isensee, David Völgyes, and Javier Coronel for their contributions. Last but not least, our special thanks go to Ludmila~I.~Kuncheva for her valuable contributions.

This work is supported by Scientific and Technological Research Council of Turkey (TUBITAK) ARDEB-EEEAG under grant number 116E133 and TUBITAK BIDEB-2214 International Doctoral Research Fellowship Programme. The work of P. Ernst, S. Chatterjee, O. Speck and, A. Nürnberger was conducted within the context of the International Graduate School MEMoRIAL at OvGU Magdeburg, Germany, supported by ESF (project no. ZS/2016/08/80646). The work of S. Aslan within the context of Ca' Foscari University of Venice is supported by under TUBITAK BIDEB-2219 grant no 1059B191701102.

\bibliographystyle{model2-names.bst}\biboptions{authoryear}
\bibliography{refs}

\end{document}